\documentclass[a4paper,12pt]{article}
\usepackage{amsfonts}
\usepackage{setspace}
\usepackage{wrapfig}
\usepackage{subfig}
\usepackage{graphicx}
\usepackage[margin=1in]{geometry}
\usepackage{natbib}
\usepackage{bbm}
\usepackage{float}
\usepackage[toc,page]{appendix}
\usepackage{color}
\usepackage{mathtools}
\usepackage{amsmath}
\usepackage{algorithm}
\usepackage{algpseudocode}
\usepackage{setspace}
\usepackage{float}

\input{tcilatex}
\begin{document}

\title{On the First Hitting Time Density of an Ornstein-Uhlenbeck Process}
\author{ Alexander Lipton\thanks{%
Massachusetts Institute of Technology, Connection Science, Cambridge, MA,
USA and Ecole Polytechnique Federale de Lausanne, Switzerland, E-mail:
alexlipt@mit.edu}, Vadim Kaushansky\thanks{%
Mathematical Institute \& Oxford-Man Institute, University of Oxford, UK,
E-mail: vadim.kaushansky@maths.ox.ac.uk}\thanks{%
The second author gratefully acknowledges support from the Economic and
Social Research Council and Bank of America Merrill Lynch}}
\date{}
\maketitle
\begin{abstract}
In this paper, we study the classical problem of the first passage hitting density of an Ornstein--Uhlenbeck process. We give two complementary (forward and backward) formulations of this problem and provide semi-analytical solutions for both. The corresponding problems are comparable in complexity. By using the method of heat potentials, we show how to reduce these problems to linear Volterra integral equations of the second kind. For small values of $t$, we solve these equations analytically by using Abel equation approximation; for larger $t$ we solve them numerically. We also provide a comparison with other known methods for finding the hitting density of interest, and argue that our method has considerable advantages and provides additional valuable insights.
\end{abstract}
\section{Introduction}

Computation of the first hitting time density of an Ornstein-Uhlenbeck process is a
long-standing problem, which still remains open.  An abstract approach applicable to this problem has been found by \cite{fortet22fonctions};  the Fortet's equation can be viewed as a variant of the Einstein-Smoluchowski equation (\cite{einstein1905molekularkinetischen} and \cite{von1906kinetischen}).  A general overview can be
found in \cite{borodin2012handbook, breiman1967first,
horowitz1985measure}. Attempts to find an analytical result have been made
since 1998 when \cite{leblanc1998path} first derived an analytical formula
which contained a mistake. Two years later, \cite{leblanc2000correction}
published a correction on the paper; unfortunately, the
correction was erroneous as well. \cite{going2003clarification} noticed that the
authors had incorrectly used a spatial homogeneity property for the
three-dimensional Bessel bridge.

\cite{alili2005representations}, \cite{linetsky2004computing}, and \cite{ricciardi1988first}
found representations for the hitting density by using the Laplace transform. \cite%
{alili2005representations} gave several representations: a representation in
the series of parabolic cylinder functions and its derivatives, an
indefinite integral representation via special functions, and a Bessel
bridge representation. The first approach is based on inverting the Laplace
transform, which is computed analytically. As a result, the authors got the series representation involving
 parabolic cylinder functions and its derivatives. The second approach is based on the cosine transform and its
inverse. The authors got a representation via an indefinite integral
involving some special functions and computed it using the trapezoidal rule.
The third approach gives a representation of the density via an expectation
of a function of the three-dimensional Bessel bridge, which can be computed
using the Monte Carlo method. \cite{linetsky2004computing} gave an analytical
representation via relevant Sturm--Liouville eigenfunction expansions. The
coefficients were found as a solution of a nonlinear equation, which
involved nonlinear special (Hermite) functions. 

\cite{martin2015infinite} solved a nonlinear Fokker--Planck equation with
steady-state solution by representing it as an infinite product rather
than -- as usual -- an infinite sum. The PDE, which corresponds to the
transition probability of Ornstein-Uhlenbeck process, belongs to the class
of equations described in \cite{martin2015infinite}. In principle, the
results of the paper can be modified to the computation of first time
hitting density. An advantage of this method is that it allows quantifying the errors; thereby controlling the number of terms required to reach a given precision.

As we show later, under a suitable change of variables, the hitting problem
of an OU process becomes the problem of hitting the square-root boundary of a
Brownian motion. \cite{novikov1971stopping} and \cite{shepp1967first}
derived the density and moments expansion for this problem; \cite%
{novikov1999approximations}, \cite{daniels1996approximating}, and \cite%
{potzelberger2001boundary} developed numerical methods for general curved
boundaries. \cite{hyer1999hidden} and \cite{avellaneda2001distance} 
analyzed the problem with curvilinear boundaries in finance.

The hitting density of an OU process has many applications in applied
mathematics, especially in mathematical finance (\cite{martin2015infinite}
for the design of trading strategies; \cite{leblanc1998path} for pricing
path-dependent options on yields; \cite{jeanblanc2000modelling},
\cite{collin2001credit}, \cite{coculescu2008valuation}, \cite{yi2010first} for credit risk modeling; and \cite%
{cheridito2015pricing} for CoCo bonds pricing). It has also applications in
quantitative biology (\cite{smith1991laguerre}), where the hitting time is
used for modeling the time between rings of a nerve cell. 

The methods described above require substantial numerical computations, and for
some of them the convergence rate is unknown. Moreover, most of them are
difficult to implement; for example, \cite{cheridito2015pricing} preferred the
Crank--Nicolson method to other known analytical methods, because it is
easier to implement.

In this paper, we develop a fast semi-analytical method to compute the first
time hitting density of an Ornstein-Uhlenbeck process, which is easy to
implement. After an appropriate change of variables, the corresponding PDE
becomes a heat equation with a moving boundary. We solve it using the method
of heat potentials (\cite{Lipton2001}, Section 12.2.3, pp. 462--467). It
leads to a Volterra equation of the second kind, which we solve numerically
similar to \cite{Lipton2018semi}. As a result, we got recursive formulas to
compute the corresponding hitting density.

The rest of the paper is organized as follows: in Section \ref{sec:prob_form}
we formulate the problem and eliminate parameters using a change of
variables; in Section \ref{sec:special_case} we solve the problem
analytically for a special case; in Section \ref{sec:general_case} we derive
the solution in terms of a Volterra equation using the method of heat
potentials; in Section \ref{sec:num_method} we consider a numerical solution
for the corresponding Volterra equation as well as a solution as an
approximation by an Abel equation; in Section \ref{sec:num_res} we show
numerical illustrations and compare the methods; in Section \ref%
{sec:conclusion} we conclude.

\section{Problem formulation and initial transformations}

\label{sec:prob_form} Consider a typical OU process%
\begin{equation*}
\begin{aligned} dX_{t} &=\lambda \left( \theta -X_{t}\right) dt+\sigma
dW_{t}, \label{Eq1} \\ X_{0} &=z. \end{aligned}
\end{equation*}%
We wish to calculate the stopping time $s=\inf \left\{ t:X_{t}\leq b\right\} 
$, where $z>b$. We introduce new variables%
\begin{equation*}
\bar{t}=\lambda t,\ \ \ \bar{X}=\frac{\sqrt{\lambda }}{\sigma }\left(
X-\theta \right) ,\ \ \ \bar{z}=\frac{\sqrt{\lambda }}{\sigma }\left(
z-\theta \right) ,\ \ \ \bar{b}=\frac{\sqrt{\lambda }}{\sigma }\left(
b-\theta \right) ,  \label{Eq2}
\end{equation*}%
and rewrite the problem as follows%
\begin{equation*}
\begin{aligned} dX_{t} &=-X_{t}dt+dW_{t}, \label{Eq3} \\ X_{0} &=z, \\ s
&=\inf \left\{ t:X_{t}\leq b\right\}, \end{aligned}
\end{equation*}%
where bars are omitted for brevity.

In this formulation we consider the hitting from above problem. But, by
symmetry, one can also consider the hitting from below. It can be formulated as
the hitting problem from above with the parameters $\tilde{z} = -z$ and $\tilde{b} = -b$.

\subsection{Forward problem}

To calculate the density of the hitting time distribution $g\left(
t,z\right) $, we need to solve the following forward problem%
\begin{equation}
\begin{aligned} p_{t}\left( t,x;z\right) &=p\left( t,x;z\right)
+xp_{x}\left( t,x;z\right) +\frac{1}{2}p_{xx}\left( t,x;z\right) ,
\label{Eq4} \\ p\left( 0,x;z\right) &=\delta \left( x-z\right), \\ p\left(
t,b;z\right) &=0. \end{aligned}
\end{equation}%
This distribution is given by%
\begin{equation}
g\left( t,z\right) =\frac{1}{2}p_{x}\left( t,b;z\right) .  \label{Eq5}
\end{equation}

\subsection{Backward problem}

Alternatively, we can solve the corresponding backward problem for the
cumulative hitting probability $G\left( t,z\right) $:%
\begin{equation}
\begin{aligned} G_{t}\left( t,z\right) &=-zG_{z}\left( t,z\right)
+\frac{1}{2}G_{zz}\left( t,z\right) , \label{Eq29} \\ G\left( 0,z\right)
&=0, \\ G\left( t,b\right) &=1. \end{aligned}
\end{equation}
Then, the first hitting density is 
\begin{equation*}
g\left( t,z\right) = G_t\left( t,z\right) .
\end{equation*}
It is clear that problem (\ref{Eq29}) is easier to solve numerically than
problem (\ref{Eq4}), whilst their analytical solutions are comparable in
complexity.

\section{Special case $b = 0$}

\label{sec:special_case} Before solving the general problem via the method
of heat potentials, let us consider a special case of $b=0$. It is well
known that Green's function for the OU process in question has the form%
\begin{equation*}
H\left( t,x;0,z\right) =\frac{\exp \left( -\left. \left( e^{t}x-z\right)
^{2}\right/ 2\eta \left( t\right) +t\right) }{\sqrt{2\pi \eta \left(
t\right) }},  \label{Eq6}
\end{equation*}%
where 
\begin{equation*}
\eta \left( t\right) =\frac{e^{2t}-1}{2}=e^{t}\sinh \left( t\right) .
\label{Eq7}
\end{equation*}%
It is clear that for this case in question, the method of images works.
Indeed, $H\left( t,x;0,-z\right) $ solves the equation in question, and so
does the difference%
\begin{equation*}
H_{0}\left( t,x;0,z\right) =H\left( t,x;0,z\right) -H\left( t,x;0,-z\right) .
\label{Eq8}
\end{equation*}%
It is easy to check that 
\begin{equation*}
H_{0}\left( 0,x;0,z\right) =\delta \left( x-z\right) ,  \label{Eq9}
\end{equation*}%
and%
\begin{equation*}
H_{0}\left( t,0;0,z\right) =\frac{\exp \left( -\left. z^{2}\right/ 2\eta
\left( t\right) +t\right) }{\sqrt{2\pi \eta \left( t\right) }}-\frac{\exp
\left( -\left. z^{2}\right/ 2\eta \left( t\right) +t\right) }{\sqrt{2\pi
\eta \left( t\right) }}=0.  \label{Eq10}
\end{equation*}%
Hence 
\begin{equation*}
p\left( t,x\right) =H_{0}\left( t,x;0,z\right) .  \label{Eq11}
\end{equation*}%
The corresponding hitting time density $q\left( t\right) $ is given by%
\begin{eqnarray}
q\left( t\right) &=&\frac{1}{2}p_{x}\left( t,0\right)  \notag \\
&=&\frac{1}{2}\left( -\frac{e^{t}\left( e^{t}x-z\right) }{\eta \left(
t\right) }\frac{\exp \left( -\left. \left( e^{t}x-z\right) ^{2}\right/ 2\eta
\left( t\right) +t\right) }{\sqrt{2\pi \eta \left( t\right) }}\right.  \notag
\\
&&\left. \left. +\frac{e^{t}\left( e^{t}x+z\right) }{\eta \left( t\right) }%
\frac{\exp \left( -\left. \left( e^{t}x+z\right) ^{2}\right/ 2\eta \left(
t\right) +t\right) }{\sqrt{2\pi \eta \left( t\right) }}\right) \right\vert
_{x=0}  \label{Eq12} \\
&=&\frac{e^{t}z}{\eta \left( t\right) }\frac{\exp \left( -\left.
z^{2}\right/ 2\eta \left( t\right) +t\right) }{\sqrt{2\pi \eta \left(
t\right) }}  \notag \\
&=&\frac{z\exp \left( -\frac{e^{-t}z^{2}}{2\sinh \left( t\right) }+\frac{t}{2%
}\right) }{\sqrt{2\pi \left( \sinh \left( t\right) \right) ^{3}}}.  \notag
\end{eqnarray}

\section{General case}

\label{sec:general_case}

\subsection{Forward problem}

Now, we consider the general case. We wish to transform IBVP (\ref{Eq4})
into the standard IBVP for a heat equation with a moving boundary. To this
end, we introduce new independent and dependent variables as follows:%
\begin{equation*}
q(\tau ,\xi )=e^{-t}p(t,x),\quad \tau =\eta \left( t\right) ,\quad \xi
=e^{t}x.  \label{Eq13}
\end{equation*}%
Given that%
\begin{eqnarray*}
\frac{\partial }{\partial t} &=&\left( 2\tau +1\right) \frac{\partial }{%
\partial \tau }+\xi \frac{\partial }{\partial \xi },  \label{Eq14} \\
x\frac{\partial }{\partial x} &=&x\sqrt{2\tau +1}\frac{\partial }{\partial
\xi }=\xi \frac{\partial }{\partial \xi },  \notag \\
\frac{\partial ^{2}}{\partial x^{2}} &=&\left( 2\tau +1\right) \frac{%
\partial ^{2}}{\partial \xi ^{2}},  \notag
\end{eqnarray*}%
we get%
\begin{equation}
\begin{aligned} q_{\tau }\left( \tau ,\xi \right) &=\frac{1}{2}q_{\xi \xi
}\left( \tau ,\xi \right) , \label{Eq15} \\ q\left( 0,\xi \right) &=\delta
\left( \xi -z\right) , \\ q\left( \tau ,b\left( \tau \right) \right) &=0, \\
b\left( \tau \right) &=\sqrt{2\tau +1}b. \end{aligned}
\end{equation}%
It is clear that $0\leq \tau <\infty $ and $e^{t}=\sqrt{2\tau +1}$. We show
the moving boundary for different values of $b$ in Figure \ref%
{moving_boundary}. 
\begin{figure}[tbp]
\begin{center}
\includegraphics[width=0.8\textwidth]{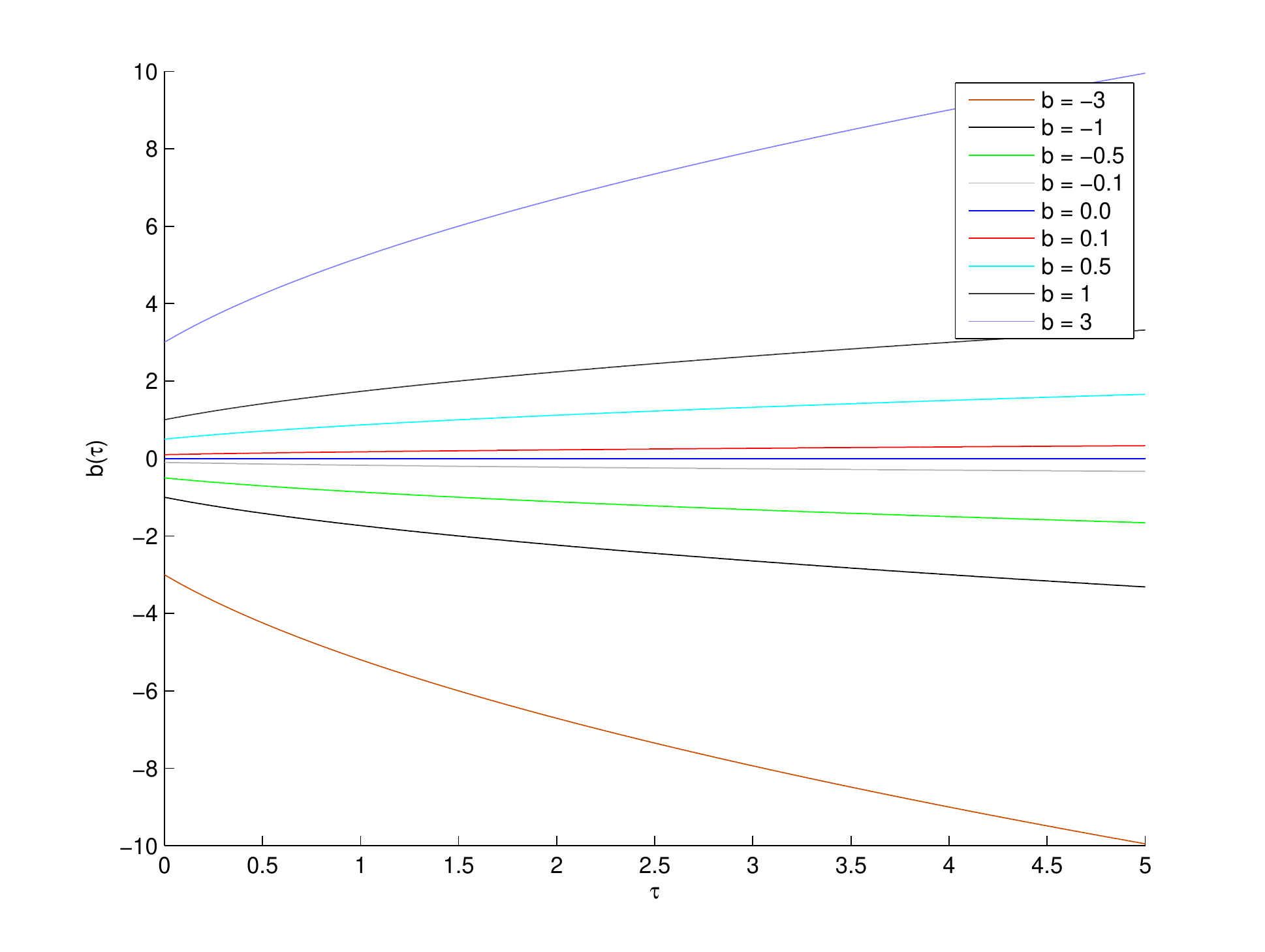}
\end{center}
\par
\vspace{-10pt}
\caption{Moving boundary $b(\protect\tau )$ for \eqref{Eq15}}
\label{moving_boundary}
\end{figure}

IBVP (\ref{Eq15}) can be solved via the method of heat potentials (see \cite%
{tikhonov2013equations}, pp. 530-535; \cite{Lipton2001}, Section 12.2.3, pp.
462-467 for details). As usual, we write%
\begin{equation*}
q\left( \tau ,\xi \right) =H\left( \tau ,\xi ,0,z\right) +\tilde{q}\left(
\tau ,\xi \right) ,  \label{Eq16}
\end{equation*}%
where $H\left( \tau ,\xi ,0,z\right) $ is the standard heat kernel, while $%
\tilde{q}\left( \tau ,\xi \right) $ solves the IBVP of the form%
\begin{eqnarray*}
\tilde{q}_{\tau }\left( \tau ,\xi \right) &=&\frac{1}{2}\tilde{q}_{\xi \xi
}\left( \tau ,\xi \right) ,  \notag \\
\tilde{q}\left( 0,\xi \right) &=&0,  \label{Eq17} \\
\tilde{q}\left( \tau ,b\left( \tau \right) \right) &=&-H\left( \tau ,b\left(
\tau \right) ,0,z\right) .  \notag
\end{eqnarray*}%
Accordingly,%
\begin{equation*}
\tilde{q}\left( \tau ,\xi \right) =\int_{0}^{\tau }\frac{\left( \xi -b\left(
\tau ^{\prime }\right) \right) \exp \left( -\frac{\left( \xi -b\left( \tau
^{\prime }\right) \right) ^{2}}{2\left( \tau -\tau ^{\prime }\right) }%
\right) }{\sqrt{2\pi \left( \tau -\tau ^{\prime }\right) ^{3}}}\nu^f \left(
\tau ^{\prime }\right) d\tau ^{\prime },  \label{Eq18}
\end{equation*}%
where $\nu^f $ is the solution of the Volterra equation of the second kind,%
\begin{equation}
\nu^f \left( \tau \right) +\int_{0}^{\tau }\frac{\left( b\left( \tau \right)
-b\left( \tau ^{\prime }\right) \right) \exp \left( -\frac{\left( b\left(
\tau \right) -b\left( \tau ^{\prime }\right) \right) ^{2}}{2\left( \tau
-\tau ^{\prime }\right) }\right) \nu^f \left( \tau ^{\prime }\right) }{\sqrt{%
2\pi \left( \tau -\tau ^{\prime }\right) ^{3}}} d\tau ^{\prime }+\frac{\exp
\left( \left( -\frac{\left( b\left( \tau \right) -z\right) ^{2}}{2\tau }%
\right) \right) }{\sqrt{2\pi \tau }}=0.  \label{Eq19}
\end{equation}

Now%
\begin{equation*}
\frac{b\left( \tau \right) -b\left( \tau ^{\prime }\right) }{\tau -\tau
^{\prime }}=\frac{b\left( \sqrt{2\tau +1}-\sqrt{2\tau ^{\prime }+1}\right) }{%
\tau -\tau ^{\prime }}=\frac{2b}{\sqrt{2\tau +1}+\sqrt{2\tau ^{\prime }+1}},
\label{Eq20}
\end{equation*}%
so that \eqref{Eq19} can be written in the form%
\begin{equation}
\nu^f \left( \tau \right) +\sqrt{\frac{2}{\pi }}b\int_{0}^{\tau }\frac{\exp
\left( -b^{2}\frac{\left( \sqrt{2\tau +1}-\sqrt{2\tau ^{\prime }+1}\right) }{%
\left( \sqrt{2\tau +1}+\sqrt{2\tau ^{\prime }+1}\right) }\right) \nu^f
\left( \tau ^{\prime }\right)}{\left( \sqrt{2\tau +1}+\sqrt{2\tau ^{\prime
}+1}\right) \sqrt{\tau -\tau ^{\prime }}} d\tau ^{\prime } +\frac{\exp
\left( -\frac{\left( \sqrt{2\tau +1}b-z\right) ^{2}}{2\tau }\right) }{\sqrt{%
2\pi \tau }}=0.  \label{Eq23}
\end{equation}%
and is easy to solve numerically.

Assuming that $\nu ^{f}\left( \tau \right) $ is found, we can proceed as
follows:%
\begin{equation}
p\left( t,x\right) =\frac{\exp \left( -\left. \left( e^{t}x-z\right)
^{2}\right/ \left( e^{2t}-1\right) +t\right) }{\sqrt{\pi \left(
e^{2t}-1\right) }}+e^{t}\tilde{q}\left( \tau ,\xi \right) ,  \label{Eq21}
\end{equation}%
Then, $g(t)$ can be computed using \eqref{Eq5} by differentiation of %
\eqref{Eq21} and taking its limit at $b$. We do the necessary computations in
Appendix \ref{appendix}, and write the final formula below: 
\begin{eqnarray}
g\left( t\right) &=&\frac{1}{2}p_{x}\left( t,b\right)  \label{Eq22} \\
&=&-\frac{\left( e^{t}b-z\right) \exp \left( -\frac{\left( e^{t}b-z\right)
^{2}}{\left( e^{2t}-1\right) }+2t\right) }{\sqrt{\pi \left( e^{2t}-1\right)
^{3}}}-\left( e^{t}b+\frac{e^{2t}}{\sqrt{\pi \left( e^{2t}-1\right) }}%
\right) \nu ^{f}\left( t\right)  \notag \\
&+&\frac{e^{2t}}{\sqrt{8\pi }}\int_{0}^{\tau }\frac{\left( 1-2b^{2}\frac{%
\left( \sqrt{2\tau +1}-\sqrt{2\tau ^{\prime }+1}\right) }{\left( \sqrt{2\tau
+1}+\sqrt{2\tau ^{\prime }+1}\right) }\right) \exp \left( -b^{2}\frac{\left( 
\sqrt{2\tau +1}-\sqrt{2\tau ^{\prime }+1}\right) }{\left( \sqrt{2\tau +1}+%
\sqrt{2\tau ^{\prime }+1}\right) }\right) \nu ^{f}\left( \tau ^{\prime
}\right) -\nu ^{f}\left( \tau \right) }{\sqrt{\left( \tau -\tau ^{\prime
}\right) ^{3}}}d\tau ^{\prime }.  \notag
\end{eqnarray}

We can rewrite (\ref{Eq23}) in an alternative way. Let $\theta =\sqrt{2\tau
+1}-1$, $\theta ^{\prime }=\sqrt{2\tau ^{\prime }+1}-1$, $0\leq \theta
^{\prime }\leq \theta <\infty $. Then%
\begin{equation}
\nu ^{f}\left( \theta \right) +\frac{2b}{\sqrt{\pi }}\int_{0}^{\theta }\frac{%
\exp \left( -b^{2}\frac{\left( \theta -\theta ^{\prime }\right) }{\left(
2+\theta +\theta ^{\prime }\right) }\right) \left( 1+\theta ^{\prime
}\right) \nu ^{f}\left( \theta ^{\prime }\right) }{\sqrt{\left( 2+\theta
+\theta ^{\prime }\right) ^{3}\left( \theta -\theta ^{\prime }\right) }}\
d\theta ^{\prime }+\frac{\exp \left( -\frac{\left( \left( 1+\theta \right)
b-z\right) ^{2}}{\left( \left( 1+\theta \right) ^{2}-1\right) }\right) }{%
\sqrt{\pi \left( \left( 1+\theta \right) ^{2}-1\right) }}=0.  \label{Eq23a}
\end{equation}%
Symbolically,%
\begin{equation*}
\nu ^{f}\left( \theta \right) +\int_{0}^{\theta }\frac{\Phi _{b}^{f}(\theta
,\theta ^{\prime })\nu ^{f}\left( \theta ^{\prime }\right) }{\sqrt{\theta
-\theta ^{\prime }}}d\theta ^{\prime }+\frac{\exp \left( -\frac{\left(
\left( 1+\theta \right) b-z\right) ^{2}}{\left( \left( 1+\theta \right)
^{2}-1\right) }\right) }{\sqrt{\pi \left( \left( 1+\theta \right)
^{2}-1\right) }}=0,
\end{equation*}%
where%
\begin{equation*}
\Phi _{b}^{f}(\theta ,\theta ^{\prime })=\frac{2b}{\sqrt{\pi }}\frac{\exp
\left( -b^{2}\frac{\left( \theta -\theta ^{\prime }\right) }{\left( 2+\theta
+\theta ^{\prime }\right) }\right) \left( 1+\theta ^{\prime }\right) }{\sqrt{%
\left( 2+\theta +\theta ^{\prime }\right) ^{3}}}.
\end{equation*}%
Accordingly, (\ref{Eq22}) can be written in the form%
\begin{eqnarray*}
g\left( t\right) &=&-\frac{\left( e^{t}b-z\right) \exp \left( -\frac{\left(
e^{t}b-z\right) ^{2}}{\left( e^{2t}-1\right) }+2t\right) }{\sqrt{\pi \left(
e^{2t}-1\right) ^{3}}}-\left( e^{t}b+\frac{e^{2t}}{\sqrt{\pi \left(
e^{2t}-1\right) }}\right) \nu ^{f}\left( t\right) \\
&+&\frac{1}{\sqrt{\pi }}%
e^{2t}  \int_{0}^{\theta }\frac{\left( \left( 1-2b^{2}\frac{\left( \theta
-\theta ^{\prime }\right) }{\left( 2+\theta +\theta ^{\prime }\right) }%
\right) \exp \left( -b^{2}\frac{\left( \theta -\theta ^{\prime }\right) }{%
\left( 2+\theta +\theta ^{\prime }\right) }\right) \nu ^{f}\left( \theta
^{\prime }\right) -\nu ^{f}\left( \theta \right) \right) \left( 1+\theta
^{\prime }\right) }{\sqrt{\left( 2+\theta +\theta ^{\prime }\right)
^{3}\left( \theta -\theta ^{\prime }\right) ^{3}}}d\theta ^{\prime }.  \notag
\end{eqnarray*}

\subsection{Backward problem}

\subsubsection{General case}

\begin{figure}[tbp]
\begin{center}
\includegraphics[width=0.8\textwidth]{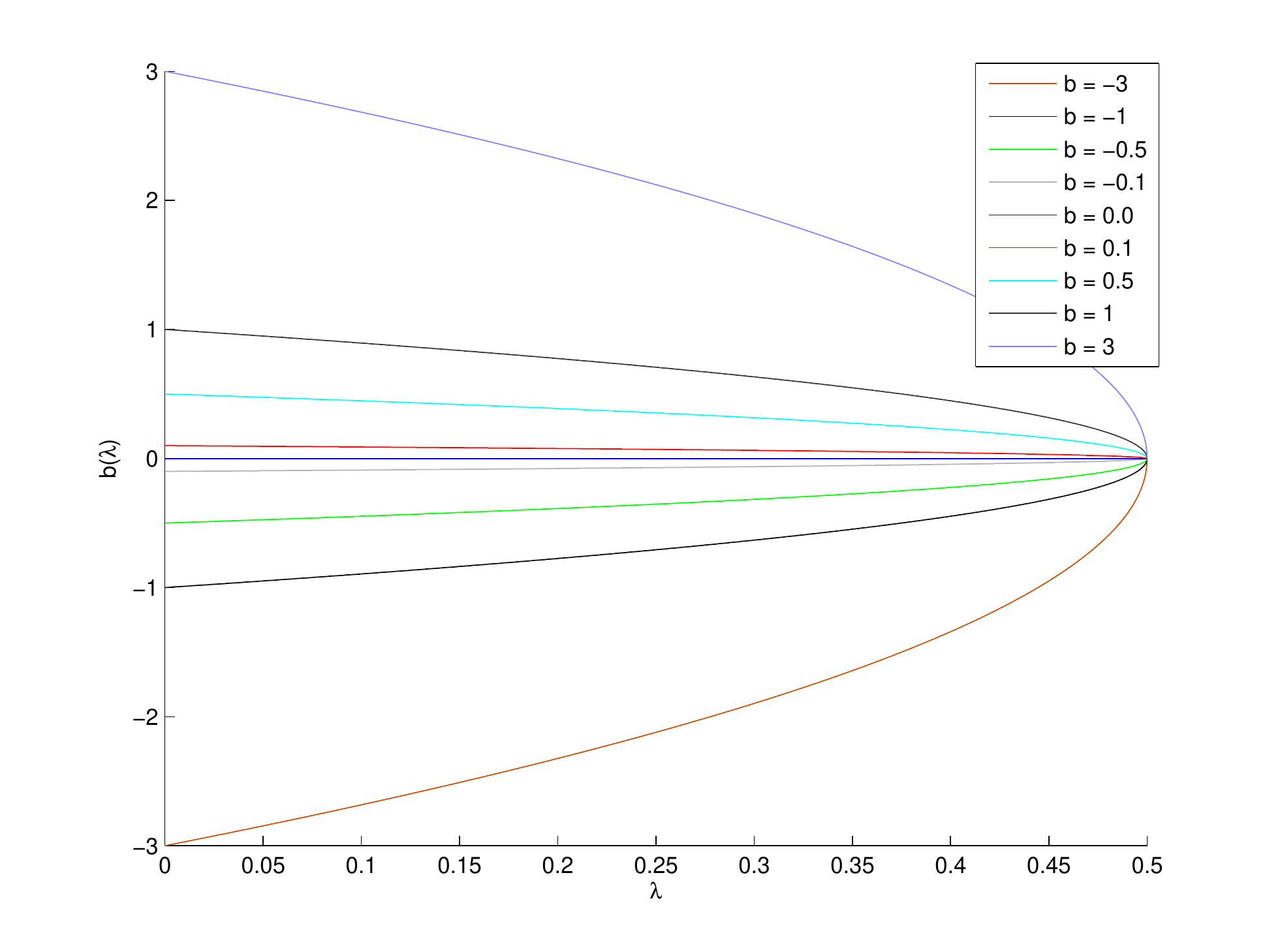}
\end{center}
\par
\vspace{-10pt}
\caption{Moving boundary $b(\protect\lambda )$ for \eqref{Eq32}}
\label{moving_boundary2}
\end{figure}

By the same token as before, we introduce%
\begin{equation}
\lambda =\varpi \left( t\right) =\frac{1-e^{-2t}}{2}=e^{-t}\sinh \left(
t\right) ,\quad 0\leq \lambda <\frac{1}{2},\quad \mu =e^{-t}z,  \label{Eq30}
\end{equation}%
notice that%
\begin{eqnarray*}
\frac{\partial }{\partial t} &=&\left( 1-2\lambda \right) \frac{\partial }{%
\partial \lambda }-\mu \frac{\partial }{\partial \mu },  \label{Eq31} \\
z\frac{\partial }{\partial z} &=&z\sqrt{1-2\lambda }\frac{\partial }{%
\partial \mu }=\mu \frac{\partial }{\partial \mu },  \notag \\
\frac{\partial ^{2}}{\partial z^{2}} &=&\left( 1-2\lambda \right) \frac{%
\partial ^{2}}{\partial \mu ^{2}},  \notag
\end{eqnarray*}%
and rewrite IBVP (\ref{Eq29}) as follows%
\begin{equation}
\begin{aligned} G_{\lambda }\left( \lambda ,\mu \right) &=\frac{1}{2}G_{\mu
\mu }\left( \lambda ,\mu \right) , \label{Eq32} \\ G\left( 0,\mu \right)
&=0, \\ G\left( \lambda ,b\left( \lambda \right) \right) &=1, \\ b\left(
\lambda \right) &=\sqrt{1-2\lambda }b. \end{aligned}
\end{equation}%
It is clear that $0\leq \lambda <1/2$, and $e^{-t}=\sqrt{1-2\lambda }$. It
is worth noting that for the backward problem the computational domain is
compactified in the $\lambda $ direction. We show the moving boundary for
different values of $b$ in Figure \ref{moving_boundary2}.

Accordingly,%
\begin{equation}
G\left( \lambda ,\mu \right) =\int_{0}^{\lambda }\frac{\left( \mu -b\left(
\lambda ^{\prime }\right) \right) \exp \left( -\frac{\left( \mu -b\left(
\lambda ^{\prime }\right) \right) ^{2}}{2\left( \lambda -\lambda ^{\prime
}\right) }\right) \nu ^{b}\left( \lambda ^{\prime }\right) }{\sqrt{2\pi
\left( \lambda -\lambda ^{\prime }\right) ^{3}}}d\lambda ^{\prime },
\label{Eq35}
\end{equation}%
\begin{eqnarray}
g\left( \lambda ,\mu \right) &=&\left( \left( 1-2\lambda \right) \frac{%
\partial }{\partial \lambda }-\mu \frac{\partial }{\partial \mu }\right)
\int_{0}^{\lambda }\frac{\left( \mu -b\left( \lambda ^{\prime }\right)
\right) \exp \left( -\frac{\left( \mu -b\left( \lambda ^{\prime }\right)
\right) ^{2}}{2\left( \lambda -\lambda ^{\prime }\right) }\right) }{\sqrt{%
2\pi \left( \lambda -\lambda ^{\prime }\right) ^{3}}}\nu ^{b}\left( \lambda
^{\prime }\right) d\lambda ^{\prime }  \label{Eq42} \\
&=&\frac{\left( 1-2\lambda \right) }{2}\int_{0}^{\lambda }\left( -3\left(
\lambda -\lambda ^{\prime }\right) +\left( \mu -b\left( \lambda ^{\prime
}\right) \right) ^{2}\right) \left( \mu -b\left( \lambda ^{\prime }\right)
\right)  \notag \\
&&\times \frac{\exp \left( -\frac{\left( \mu -b\left( \lambda ^{\prime
}\right) \right) ^{2}}{2\left( \lambda -\lambda ^{\prime }\right) }\right) }{%
\sqrt{2\pi \left( \lambda -\lambda ^{\prime }\right) ^{7}}}\nu ^{b}\left(
\lambda ^{\prime }\right) d\lambda ^{\prime }  \notag \\
&&+\mu \int_{0}^{\lambda }\left( -\left( \lambda -\lambda ^{\prime }\right)
+\left( \mu -b\left( \lambda ^{\prime }\right) \right) ^{2}\right) \frac{%
\exp \left( -\frac{\left( \mu -b\left( \lambda ^{\prime }\right) \right) ^{2}%
}{2\left( \lambda -\lambda ^{\prime }\right) }\right) }{\sqrt{2\pi \left(
\lambda -\lambda ^{\prime }\right) ^{5}}}\nu ^{b}\left( \lambda ^{\prime
}\right) d\lambda ^{\prime }.  \notag
\end{eqnarray}%
where $\nu ^{b}$ is the solution of the Volterra equation of the second kind,%
\begin{equation*}
\nu ^{b}\left( \lambda \right) +\int_{0}^{\lambda }\frac{\left( b\left(
\lambda \right) -b\left( \lambda ^{\prime }\right) \right) \exp \left( -%
\frac{\left( b\left( \lambda \right) -b\left( \lambda ^{\prime }\right)
\right) ^{2}}{2\left( \lambda -\lambda ^{\prime }\right) }\right) \nu
^{b}\left( \lambda ^{\prime }\right) }{\sqrt{2\pi \left( \lambda -\lambda
^{\prime }\right) ^{3}}}d\lambda ^{\prime }-1=0.  \label{Eq36}
\end{equation*}

Since%
\begin{equation}
\frac{b\left( \lambda \right) -b\left( \lambda ^{\prime }\right) }{\lambda
-\lambda ^{\prime }}=\frac{b\left( \sqrt{1-2\lambda }-\sqrt{1-2\lambda
^{\prime }}\right) }{\lambda -\lambda ^{\prime }}=-\frac{2b}{\sqrt{%
1-2\lambda }+\sqrt{1-2\lambda ^{\prime }}},  \label{Eq37}
\end{equation}%
so that \eqref{Eq37} can be written in the form%
\begin{equation}
\nu ^{b}\left( \lambda \right) -\sqrt{\frac{2}{\pi }}b\int_{0}^{\lambda }%
\frac{\exp \left( -b^{2}\frac{\left( \sqrt{1-2\lambda ^{\prime }}-\sqrt{%
1-2\lambda }\right) }{\left( \sqrt{1-2\lambda ^{\prime }}+\sqrt{1-2\lambda }%
\right) }\right) \nu ^{b}\left( \lambda ^{\prime }\right) }{\left( \sqrt{%
1-2\lambda ^{\prime }}+\sqrt{1-2\lambda }\right) \sqrt{\lambda -\lambda
^{\prime }}}d\lambda ^{\prime }-1=0.  \label{Eq38}
\end{equation}%
Once \eqref{Eq38} is solved, $g\left( \lambda ,\mu \right) $ and $g\left(
t,z\right) $ can be calculated by virtue of \eqref{Eq35} and \eqref{Eq30} in
a straightforward fashion.

We can rewrite (\ref{Eq38}) in an alternative way. Let $\vartheta =1-\sqrt{%
1-2\lambda }$, $\vartheta ^{\prime }=1-\sqrt{1-2\lambda ^{\prime }}$, $0\leq
\vartheta ^{\prime }\leq \vartheta <1$. Then 
\begin{equation}
\nu ^{b}\left( \vartheta \right) -\frac{2b}{\sqrt{\pi }}\int_{0}^{\vartheta }%
\frac{\exp \left( -b^{2}\frac{\left( \vartheta -\vartheta ^{\prime }\right) 
}{\left( 2-\vartheta -\vartheta ^{\prime }\right) }\right) \left(
1-\vartheta ^{\prime }\right) \nu ^{b}\left( \vartheta ^{\prime }\right) }{%
\sqrt{\left( 2-\vartheta -\vartheta ^{\prime }\right) ^{3}}\sqrt{\vartheta
-\vartheta ^{\prime }}}d\vartheta ^{\prime }-1=0.  \label{Eq38a}
\end{equation}%
Symbolically,%
\begin{equation*}
\nu ^{b}\left( \vartheta \right) -\int_{0}^{\vartheta }\frac{\Phi
_{b}^{b}(\vartheta ,\vartheta ^{\prime })\nu ^{b}\left( \vartheta ^{\prime
}\right) }{\sqrt{\vartheta -\vartheta ^{\prime }}}d\vartheta ^{\prime }-1=0,
\end{equation*}%
where%
\begin{equation*}
\Phi _{b}^{b}(\vartheta ,\vartheta ^{\prime })=\frac{2b}{\sqrt{\pi }}\frac{%
\exp \left( -b^{2}\frac{\left( \vartheta -\vartheta ^{\prime }\right) }{%
\left( 2-\vartheta -\vartheta ^{\prime }\right) }\right) \left( 1-\vartheta
^{\prime }\right) }{\sqrt{\left( 2-\vartheta -\vartheta ^{\prime }\right)
^{3}}}.
\end{equation*}%
As one would expect, 
\begin{equation*}
\Phi _{b}^{b}(\theta ,\theta ^{\prime })=-i\Phi _{ib}^{f}\left( -\theta
,-\theta ^{\prime }\right) .
\end{equation*}%
As a result, (\ref{Eq35}), (\ref{Eq42}) become%
\begin{equation*}
G\left( t,z\right) =2\int_{0}^{\vartheta }\frac{\left( z\left( 1-\vartheta
\right) -b\left( 1-\vartheta ^{\prime }\right) \right) \exp \left( -\frac{%
\left( z\left( 1-\vartheta \right) -b\left( 1-\vartheta ^{\prime }\right)
\right) ^{2}}{\left( \vartheta -\vartheta ^{\prime }\right) \left(
2-\vartheta -\vartheta ^{\prime }\right) }\right) \left( 1-\vartheta
^{\prime }\right) \nu ^{b}\left( \vartheta ^{\prime }\right) }{\sqrt{\pi
\left( \vartheta -\vartheta ^{\prime }\right) ^{3}\left( 2-\vartheta
-\vartheta ^{\prime }\right) ^{3}}}d\vartheta ^{\prime },
\end{equation*}%
\begin{eqnarray*}
g\left( t,z\right) &=&4e^{-2t}\int_{0}^{\vartheta }\left( -3\left( \vartheta
-\vartheta ^{\prime }\right) \left( 2-\vartheta -\vartheta ^{\prime }\right)
+\left( z\left( 1-\vartheta \right) -b\left( 1-\vartheta ^{\prime }\right)
\right) ^{2}\right) \\
&&\times \left( z\left( 1-\vartheta \right) -b\left( 1-\vartheta ^{\prime
}\right) \right) \frac{\exp \left( -\frac{\left( z\left( 1-\vartheta \right)
-b\left( 1-\vartheta ^{\prime }\right) \right) ^{2}}{\left( \vartheta
-\vartheta ^{\prime }\right) \left( 2-\vartheta -\vartheta ^{\prime }\right) 
}\right) \left( 1-\vartheta ^{\prime }\right) \nu ^{b}\left( \vartheta
^{\prime }\right) }{\sqrt{\pi \left( \vartheta -\vartheta ^{\prime }\right)
^{7}\left( 2-\vartheta -\vartheta ^{\prime }\right) ^{7}}}d\vartheta
^{\prime }  \notag \\
&&+4e^{-t}z\int_{0}^{\vartheta }\left( -\left( \vartheta -\vartheta ^{\prime
}\right) \left( 2-\vartheta -\vartheta ^{\prime }\right) +\left( z\left(
1-\vartheta \right) -b\left( 1-\vartheta ^{\prime }\right) \right)
^{2}\right)  \notag \\
&&\times \frac{\exp \left( -\frac{\left( z\left( 1-\vartheta \right)
-b\left( 1-\vartheta ^{\prime }\right) \right) ^{2}}{\left( \vartheta
-\vartheta ^{\prime }\right) \left( 2-\vartheta -\vartheta ^{\prime }\right) 
}\right) \left( 1-\vartheta ^{\prime }\right) \nu ^{b}\left( \vartheta
^{\prime }\right) }{\sqrt{\pi \left( \vartheta -\vartheta ^{\prime }\right)
^{5}\left( 2-\vartheta -\vartheta ^{\prime }\right) ^{5}}}d\vartheta
^{\prime },  \notag
\end{eqnarray*}%
where $\vartheta =1-e^{-t}$.

\subsubsection{Special case}

When $b=0$, we have%
\begin{equation*}
\nu ^{b}\left( \lambda \right) =1,  \label{Eq39}
\end{equation*}%
\begin{eqnarray*}
G\left( \lambda ,\mu \right) &=&\mu \int_{0}^{\lambda }\frac{\exp \left( -%
\frac{^{\mu 2}}{2\left( \lambda -\lambda ^{\prime }\right) }\right) }{\sqrt{%
2\pi \left( \lambda -\lambda ^{\prime }\right) ^{3}}}d\lambda ^{\prime } 
\notag \\
&=&\mu \int_{0}^{\lambda }\frac{\exp \left( -\frac{^{\mu 2}}{2\lambda
^{\prime }}\right) }{\sqrt{2\pi \lambda ^{\prime 3}}}d\lambda ^{\prime } 
\notag \\
&=&2\mu \int_{1/\sqrt{\lambda }}^{\infty }\frac{\exp \left( -\frac{^{\mu
2}u^{2}}{2}\right) }{\sqrt{2\pi }}du  \label{Eq39a} \\
&=&2\int_{\mu /\sqrt{\lambda }}^{\infty }\frac{\exp \left( -\frac{v^{2}}{2}%
\right) }{\sqrt{2\pi }}dv  \notag \\
&=&2N\left( -\frac{\mu }{\sqrt{\lambda }}\right) .  \notag
\end{eqnarray*}%
Thus,%
\begin{eqnarray*}
g\left( \lambda ,\mu \right) &=&\left( \left( 1-2\lambda \right) \frac{%
\partial }{\partial \lambda }-\mu \frac{\partial }{\partial \mu }\right)
G\left( \lambda ,\mu \right)  \notag \\
&=&2\left( \left( 1-2\lambda \right) \frac{\mu }{2\sqrt{\lambda ^{3}}}+\frac{%
\mu }{\sqrt{\lambda }}\right) \frac{\exp \left( -\frac{\mu ^{2}}{2\lambda }%
\right) }{\sqrt{2\pi }}  \notag \\
&=&\frac{\mu }{\sqrt{\lambda ^{3}}}\left( 1-2\lambda +2\lambda \right) \frac{%
\exp \left( -\frac{\mu ^{2}}{2\lambda }\right) }{\sqrt{2\pi }}  \label{Eq39b}
\\
&=&\frac{\mu }{\sqrt{\lambda ^{3}}}\frac{\exp \left( -\frac{\mu ^{2}}{%
2\lambda }\right) }{\sqrt{2\pi }},  \notag
\end{eqnarray*}%
\begin{equation*}
g\left( t,z\right) =\frac{z\exp \left( -\frac{e^{-t}z^{2}}{2\sinh \left(
t\right) }+\frac{t}{2}\right) }{\sqrt{2\pi \left( \sinh \left( t\right)
\right) ^{3}}}.  \label{Eq39c}
\end{equation*}%
Alternatively,%
\begin{equation}
G\left( t,z\right) =2N\left( -\frac{e^{-t/2}z}{\sqrt{\sinh \left( t\right) }}%
\right) ,  \label{Eq40}
\end{equation}%
and%
\begin{eqnarray*}
g\left( t,z\right) &=&G_{t}\left( t,z\right)  \notag \\
&=&2\left( \frac{e^{-t/2}z}{2\sqrt{\sinh \left( t\right) }}+\frac{\cosh
\left( t\right) e^{-t/2}z}{2\sqrt{\left( \sinh \left( t\right) \right) ^{3}}}%
\right) \frac{\exp \left( -\frac{e^{-t}z^{2}}{2\sinh \left( t\right) }%
\right) }{\sqrt{2\pi }}  \label{Eq41} \\
&=&\frac{e^{-t/2}z\left( \sinh \left( t\right) +\cosh \left( t\right)
\right) }{\sqrt{\left( \sinh \left( t\right) \right) ^{3}}}\frac{\exp \left(
-\frac{e^{-t}z^{2}}{2\sinh \left( t\right) }\right) }{\sqrt{2\pi }}  \notag
\\
&=&\frac{z\exp \left( -\frac{e^{-t}z^{2}}{2\sinh \left( t\right) }+\frac{t}{2%
}\right) }{\sqrt{2\pi \left( \sinh \left( t\right) \right) ^{3}}},  \notag
\end{eqnarray*}%
as before.


\section{Numerical solution}

\label{sec:num_method} In this section, we show how to solve the Volterra
equations \eqref{Eq23a} and \eqref{Eq38a} numerically, and also derive an analytical
approximation for small values of $t$.

\subsection{Numerical method}

\label{section:num_method}

In this section, we briefly discuss two methods to solve the corresponding
Volterra equations \eqref{Eq23a} and \eqref{Eq38a}. We start with a simple
trapezoidal method, and then consider a more advanced method based on a
quadratic interpolation, which gives a better convergence rate. Both methods
can be found in \cite{linz1985analytical} (Section 8.2 and Section 8.4).

In this section we solve 
\begin{equation}  \label{volterra_gen}
f(t) = g(t) + \int_0^t \frac{K(t, s)}{\sqrt{t-s}} f(s) \, ds,
\end{equation}
where $K(t, s)$ is a non-singular part of the kernel.

Both \eqref{Eq23a} and \eqref{Eq38a} can be formulated as %
\eqref{volterra_gen} with an appropriate choice of $K(t, s)$ and $g(t)$.

For the forward equation we take (in $(\theta, \theta^{\prime })$ variables) 
\begin{equation*}
g(\theta) = -\frac{\exp \left( -\frac{\left( \left( 1+\theta \right)
b-z\right) ^{2}}{\left( \left( 1+\theta \right) ^{2}-1\right) }\right) }{%
\sqrt{\pi \left( \left( 1+\theta \right) ^{2}-1\right) }}, \quad K(\theta,
\theta^{\prime }) = -\frac{2b}{\sqrt{\pi }}\frac{\exp \left( -b^{2}\frac{%
\left( \theta -\theta ^{\prime }\right) }{\left( 2+\theta +\theta ^{\prime
}\right) }\right) \left( 1+\theta ^{\prime }\right) }{\sqrt{\left( 2+\theta
+\theta ^{\prime }\right) ^{3} }},
\end{equation*}
and for the backward equation we take (in $(\vartheta, \vartheta^{\prime })$
variables) 
\begin{equation*}
g(\vartheta) = 1, \quad K(\vartheta, \vartheta^{\prime }) = \frac{2b}{\sqrt{%
\pi }} \frac{\exp \left( -b^{2}\frac{\left( \vartheta -\vartheta ^{\prime
}\right) }{\left( 2-\vartheta -\vartheta ^{\prime }\right) }\right) \left(
1-\vartheta ^{\prime }\right) }{\sqrt{\left( 2-\vartheta -\vartheta ^{\prime
}\right) ^{3}}}.
\end{equation*}

\subsubsection{Simple trapezoidal method}

Consider the integral in \eqref{volterra_gen} separately 
\begin{equation}  \label{int_trans}
\int_{0}^{t}\frac{ K(t ,s)\nu \left(s\right) }{\sqrt{t-s }}d s=-2\int_{0}^{t
}K (t, s)\nu \left( s\right) \,d\sqrt{t-s}.
\end{equation}%
We consider a grid $0=t_0 < t_1 < \ldots < t_N = T$, and denote $F_k$ for
the approximated value of $f(t_k)$ and $\Delta _{k,l}=t_{k}-t _{l}$. Then,
by trapezoidal rule of the Stieltjes integral \eqref{int_trans}, %
\eqref{volterra_gen} can be approximated as

\begin{equation*}
F_k = g(t_k) + \sum_{i=1}^{k}\left( K(t_k, t_i) F_i +K(t_k, t_{i-1})
F_{i-1}\right) \left( \sqrt{\Delta _{k,i-1}}-\sqrt{\Delta _{k,i}}\right) =0.
\end{equation*}%
From the last equation we can express $F_k$ 
\begin{multline*}
F_k = \left( 1- K(t_k, t_k)\sqrt{\Delta _{k,k-1}}\right)^{-1} \\
\times \left( g(t_k) + K(t_k, 0) \left( \sqrt{\Delta _{k,0}}-\sqrt{\Delta
_{k,1}}\right) +\sum_{i=1}^{k-1}K(t_k, t_i) \left( \sqrt{\Delta _{k,i-1}}-%
\sqrt{\Delta _{k,i+1}}\right) F_i\right).
\end{multline*}
Taking $F_0 = g(0)$, we can recursively compute $F_m$ using the previous
values $F_0, \ldots, F_{m-1}$.

The approximation error of the integrals is of order $O(\Delta^2)$, where $%
\Delta = \max_{k, i} \sqrt{\Delta_{k, i-1}} - \sqrt{\Delta_{k, i+1}} $ is
the step size. Hence, on the uniform grid $t_i = i \Delta$, the convergence
rate is of order $O(h)$.

\subsubsection{Block-by-block method based on quadratic interpolation}

Now we consider a block-by-block method based on quadratic interpolation
from \cite{linz1985analytical} (Section 8.4).

Using piece-wise quadratic interpolation, \cite{linz1985analytical} derived 
\begin{equation}  \label{F_2m1_eq}
\begin{aligned} F_{2m+1} = g(t_{2m+1}) &+ (1-\delta_{m0}) \sum_{i = 0}^{2m}
w_{2m+1, i} K(t_{2m+1}, t_i) F_i \\ &+ \alpha \left(t_{2m+1}, t_{2m},
\frac{h}{2} \right) K(t_{2m+1}, t_{2m}) F_{2m} \\ &+ \beta \left(t_{2m+1},
t_{2m}, \frac{h}{2} \right) K(t_{2m+1}, t_{2m+1/2}) \left(\frac{3}{8} F_{2m}
+ \frac{3}{4} F_{2m+1} - \frac{1}{8} F_{2m+2} \right)\\ &+\gamma
\left(t_{2m+1}, t_{2m}, \frac{h}{2} \right) K(t_{2m+1}, t_{2m+1}) F_{2m+1},
\end{aligned}
\end{equation}
and 
\begin{equation}  \label{F_2m2_eq}
\begin{aligned} F_{2m+2} = g(t_{2m+2}) &+ (1-\delta_{m0}) \sum_{i = 0}^{2m}
w_{2m+2, i} K(t_{2m+2}, t_i) F_i \\ &+ \alpha \left(t_{2m+2}, t_{2m}, h
\right) K(t_{2m+2}, t_{2m}) F_{2m} \\ &+ \beta \left(t_{2m+2}, t_{2m}, h
\right) K(t_{2m+2}, t_{2m+1}) F_{2m+1}\\ &+\gamma \left(t_{2m+2}, t_{2m}, h
\right) K(t_{2m+2}, t_{2m+2}) F_{2m+2}, \end{aligned}
\end{equation}
where 
\begin{align*}
w_{n, i} &= (1 - \delta_{i,n-1}) \alpha(t_n, t_i, h) + (1-\delta_{i, 0})
\gamma(t_n, t_i-2h, h), & i \text{ is even}, \\
w_{n, i} &= \beta(t_n, t_i-h, h), & i \text{ is odd},
\end{align*}
and $\delta_{ij}$ is a Kronecker delta.

The functions $\alpha, \beta$, and $\gamma$ are defined by 
\begin{align*}
\alpha(x, y, z) &= \frac{z}{2} \int_0^2 \frac{(1-s)(2-s)}{\sqrt{x-y-sz}} \,
ds, \\
\beta(x, y, z) &= z \int_0^2 \frac{s(2-s)}{\sqrt{x-y-sz}} \, ds, \\
\gamma(x, y, z) &= \frac{z}{2} \int_0^2 \frac{s(s-1)}{\sqrt{x-y-sz}} \, ds.
\end{align*}
We note that $\alpha, \beta$, and $\gamma$ depend only on $x-y, z$, and can be written as a function of two variables. Moreover, these integrals can be computed analytically.

Then, one can find $[F_{2m+1}, F_{2m+2}]$ by solving the system of two
linear equations \eqref{F_2m1_eq} and \eqref{F_2m2_eq}. We present the
numerical algorithm in Algorithm \ref{algo1}. 
\begin{algorithm}[H]
	\caption{Block-by-block method based on quadratic interpolation}
	\begin{algorithmic}[1]
		\Require{$N$ --- number of time steps: $0 = t_0 < t_1 < t_2 < \ldots < t_N = T$}
		\State{$F_0 = f(0)$}
		\For{$m = 0:N/2$}
			\State{Compute $w_{2m+1, i}$ for $i = 0,\ldots, 2m$}
			\State{Compute $w_{2m+2, i}$ for $i = 0,\ldots, 2m$}
			\State{Get $[F_{2m+1}, F_{2m+2}]$ by solving  \eqref{F_2m1_eq} and \eqref{F_2m2_eq}}
		\EndFor
	\end{algorithmic}
	\label{algo1}
\end{algorithm}

Under some assumptions on regularity, the convergence rate of this method is 
$3.5$. In our case, for the backward equation, these assumptions are not satisfied, and we empirically
confirm the convergence of order $1.5$. 

\subsection{Approximation by an Abel integral equation}

For small values of $\theta $, (\ref{Eq23}) can be approximated by an Abel
integral equation of the second kind.%
\begin{equation}
\nu^f \left( \theta \right) + \frac{b}{\sqrt{2\pi }} \int_{0}^{\theta }\frac{%
1}{\sqrt{\theta -\theta ^{\prime }}}\nu^f \left( \theta ^{\prime }\right)
d\theta ^{\prime }+\frac{\exp \left( -\frac{\left( b-z\right) ^{2}}{2\theta }%
\right) }{\sqrt{2\pi \theta }}=0.  \label{abel_eq_f}
\end{equation}%
The last equation is an Abel equation of the second kind and can be solved
analytically using direct and inverse Laplace transforms (\cite{Abramowitz}).

The Laplace transform yields 
\begin{equation*}
\bar{\nu}^{f}(\Lambda )+b\frac{\bar{\nu}^{f}(\Lambda )}{\sqrt{2\Lambda }}+%
\frac{e^{-\sqrt{2\Lambda }(z-b)}}{\sqrt{2\Lambda }}=0.
\end{equation*}%
Then, $\bar{\nu}^{f}(\Lambda )$ can be expressed as 
\begin{equation*}
\bar{\nu}^{f}(\Lambda )=-\frac{e^{-\sqrt{2\Lambda }(z-b)}}{\sqrt{2\Lambda }+b%
}
\end{equation*}%
Taking inverse Laplace transform, we get the final expression for $\nu
^{f}(\theta )$ 
\begin{equation}
\nu ^{f}(\theta )=be^{\frac{b^{2}}{2}\theta +b(z-b)}N\left( -\frac{b\theta
+z-b}{\sqrt{\theta }}\right) -\frac{\exp \left( -\frac{\left( b-z\right) ^{2}%
}{2\theta }\right) }{\sqrt{2\pi \theta }},  \label{abel_analytical_f}
\end{equation}%
where $N(x)$ is the CDF of the standard normal distribution.

Now consider the backward equation. For small values of $\vartheta $, %
\eqref{Eq38a} can be approximated by 
\begin{equation}
\nu ^{b}(\vartheta )-\frac{b}{\sqrt{2\pi }}\int_{0}^{\vartheta }\frac{1}{%
\sqrt{\vartheta -\vartheta ^{\prime }}}\nu ^{b}(\vartheta ^{\prime
})\,d\vartheta ^{\prime }-1=0.  \label{abel_eq}
\end{equation}%
Similar to the forward equation, we solve it by taking direct and inverse
Laplace transforms. The direct Laplace transform yields 
\begin{equation*}
\bar{\nu}^{b}(\Lambda )-b\frac{\bar{\nu}^{b}(\Lambda )}{\sqrt{2\Lambda }}-%
\frac{1}{\Lambda }=0.
\end{equation*}%
Hence, 
\begin{equation*}
\bar{\nu}^{b}(\Lambda )=\frac{1}{\sqrt{\Lambda }(\sqrt{\Lambda }-b/\sqrt{2})}%
.
\end{equation*}%
Taking the inverse Laplace transform, we get 
\begin{equation}
\nu ^{b}(\vartheta )=2e^{\frac{b^{2}\vartheta }{2}}N(b\sqrt{\vartheta }).
\label{abel_analytical}
\end{equation}

Alternatively, one can find an analytical solution using the following
results (\cite{polyanin1998handbook}). The solution of Abel equation 
\begin{equation*}
y(t) + \xi \int_0^t \frac{y(s) ds}{\sqrt{t-s}} = f(t).
\end{equation*}
has the form 
\begin{equation*}
y(t) = F(t) + \pi \xi^2 \int_0^t \exp[\pi \xi^2 (t-s)] F(s) \, ds,
\end{equation*}
where 
\begin{equation*}
F(t) = f(t) - \xi \int_0^t \frac{f(s) \, ds}{\sqrt{t-s}}.
\end{equation*}

\section{Numerical examples}

\label{sec:num_res}

\subsection{Numerical tests}

\begin{figure}[]
\begin{center}
\subfloat[]{\includegraphics[width=0.5\textwidth]{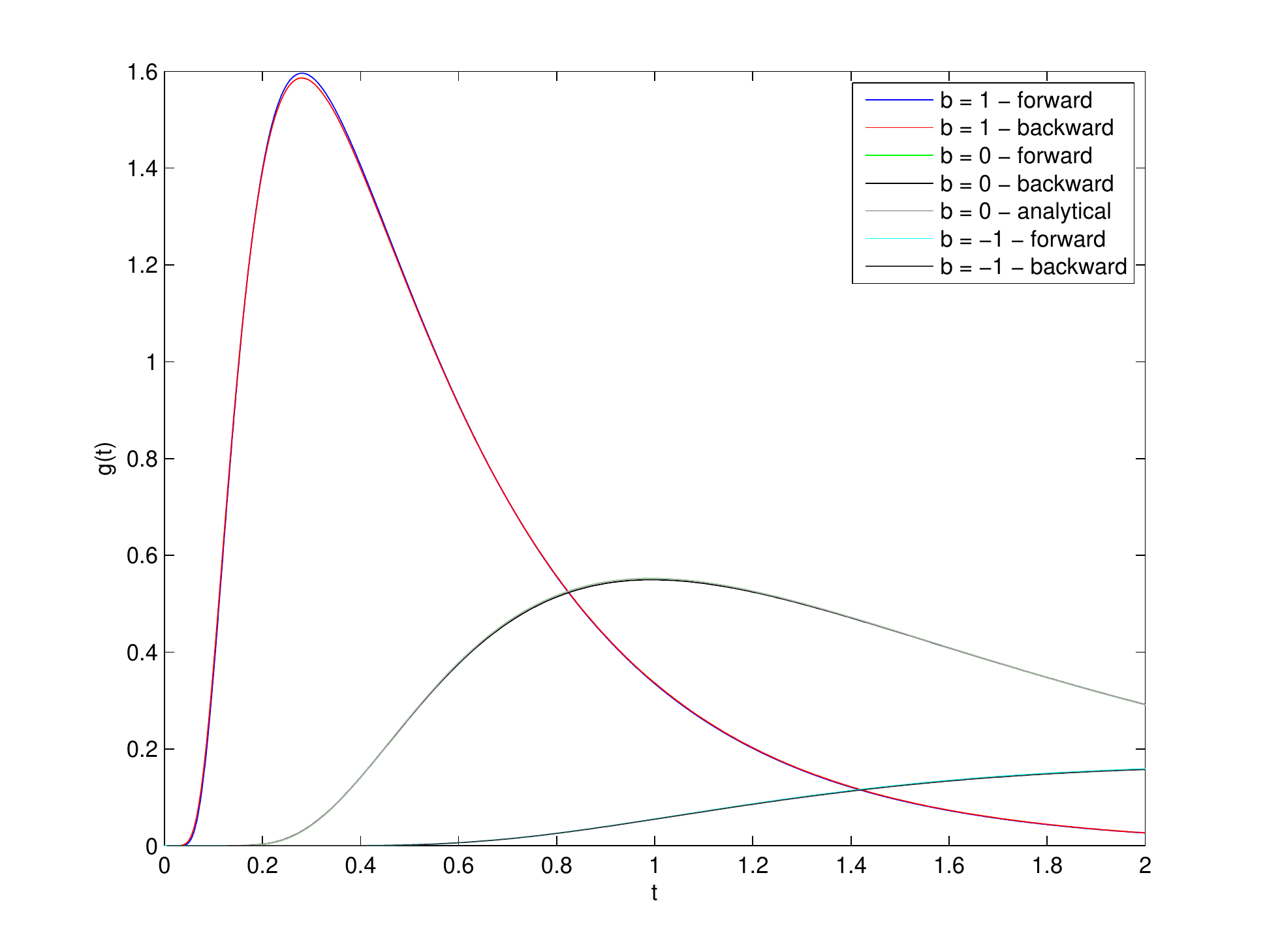}} %
\subfloat[]{\includegraphics[width=0.5\textwidth]{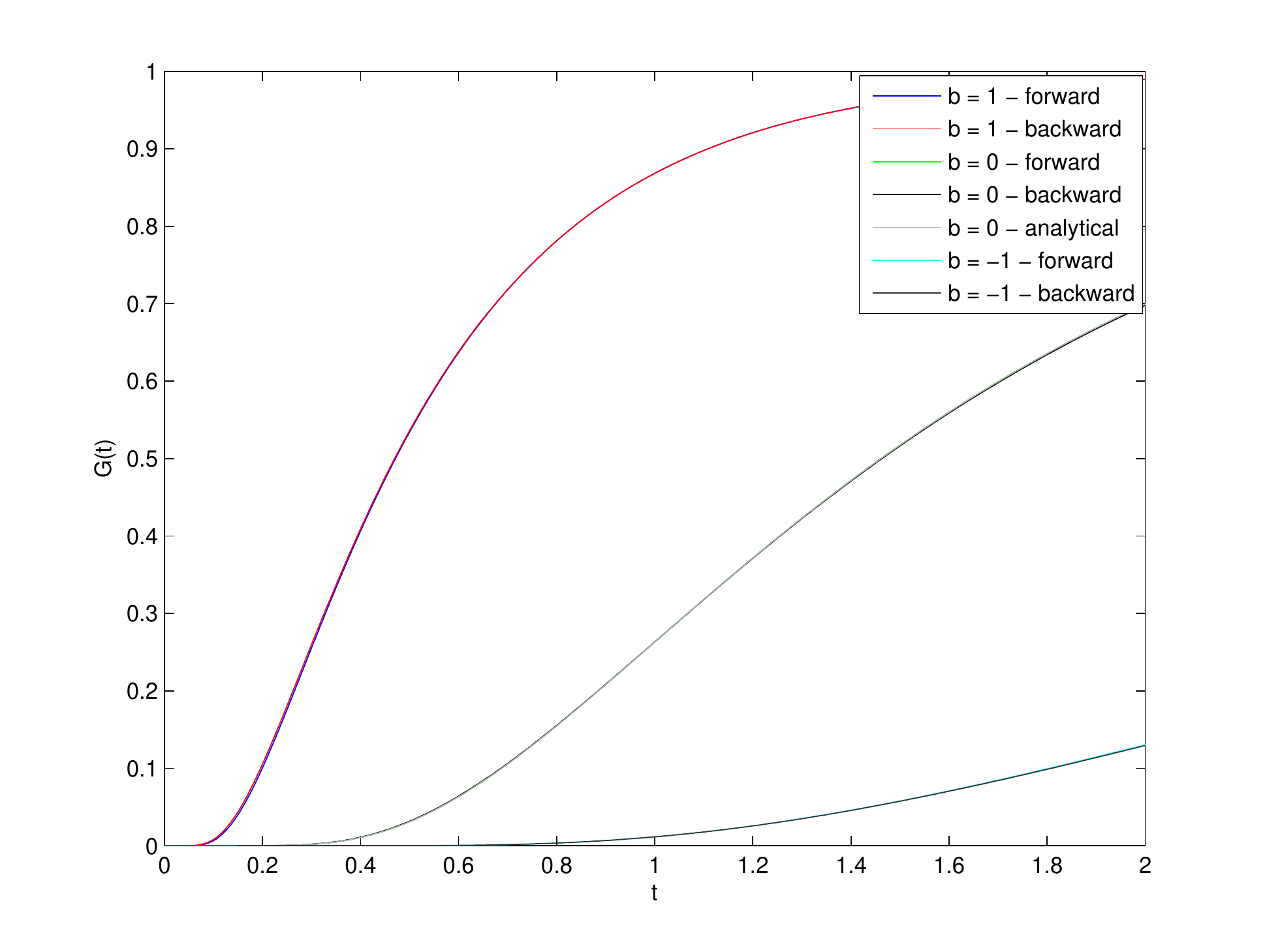}}
\end{center}
\par
\vspace{-10pt}
\caption{ (a) Density function $g(t)$ (b) Cumulative density function $G(t)$%
. }
\label{num_ex1}
\end{figure}

\begin{figure}[]
\begin{center}
\includegraphics[width=0.55\textwidth]{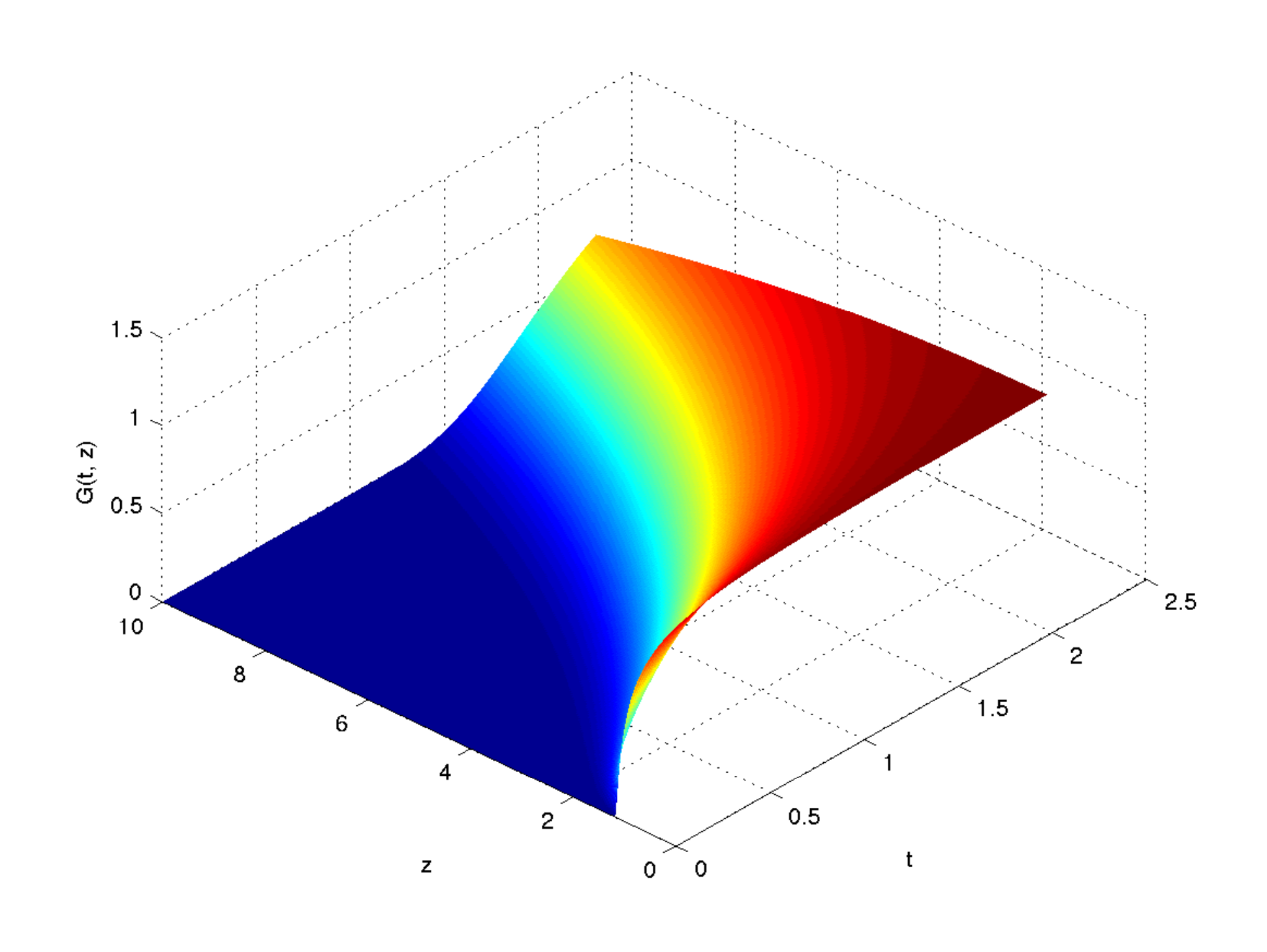}  %
\end{center}
\par
\vspace{-10pt}
\caption{ $G(t, z)$ as a function of both $t$ and $z$ for $b = 1$.%
 }
 \label{G_t_z}
\end{figure}

\begin{figure}[]
\begin{center}
\subfloat[]{\includegraphics[width=0.5\textwidth]{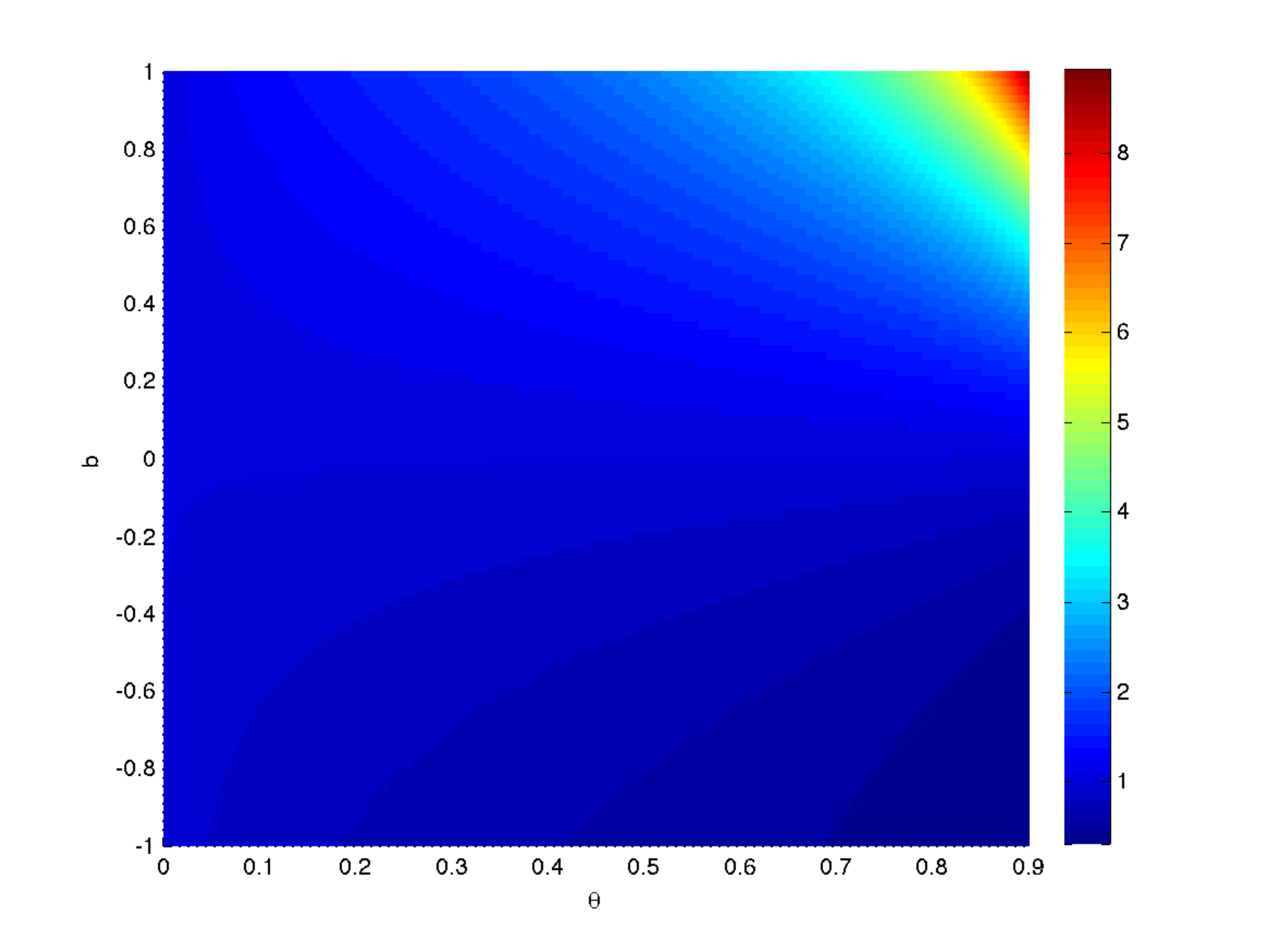}} %
\subfloat[]{\includegraphics[width=0.5\textwidth]{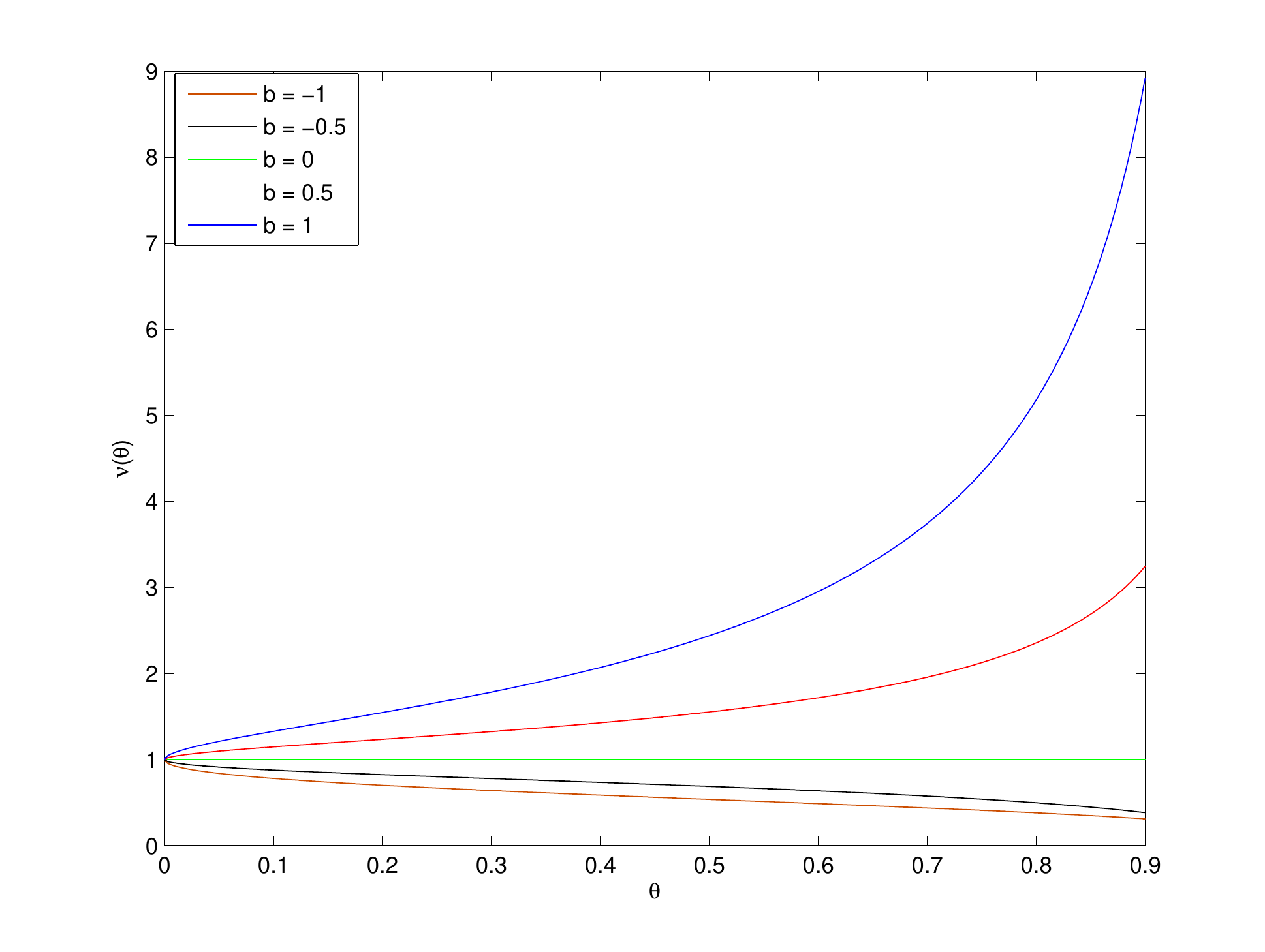}}
\end{center}
\par
\vspace{-10pt}
\caption{ $\protect\nu^b(\protect\vartheta)$, the solution of \eqref{Eq38a}
(a) as a function of $\protect\vartheta$ and $b$ (b) as a function of $%
\protect\vartheta$ for different values of $b$.}
\label{nu_backward}
\end{figure}

Consider $T = 2$ and $z = 2$. In Figure \ref{num_ex1} we examine the density
and cumulative density functions for different values of $b$ and compare them
with the analytical solution \eqref{Eq12} for $b = 0$. In Figure \ref{G_t_z} we show $G(t, z)$ 
as a function of both $t$ and $z$ for $b = 1$. In Figure \ref%
{nu_forward} we present $\nu^f(\tau)$ and in Figure \ref{nu_backward} we
present $\nu^b(\vartheta)$ for various values of $b$. 

\begin{figure}[]
\begin{center}
\subfloat[]{\includegraphics[width=0.5\textwidth]{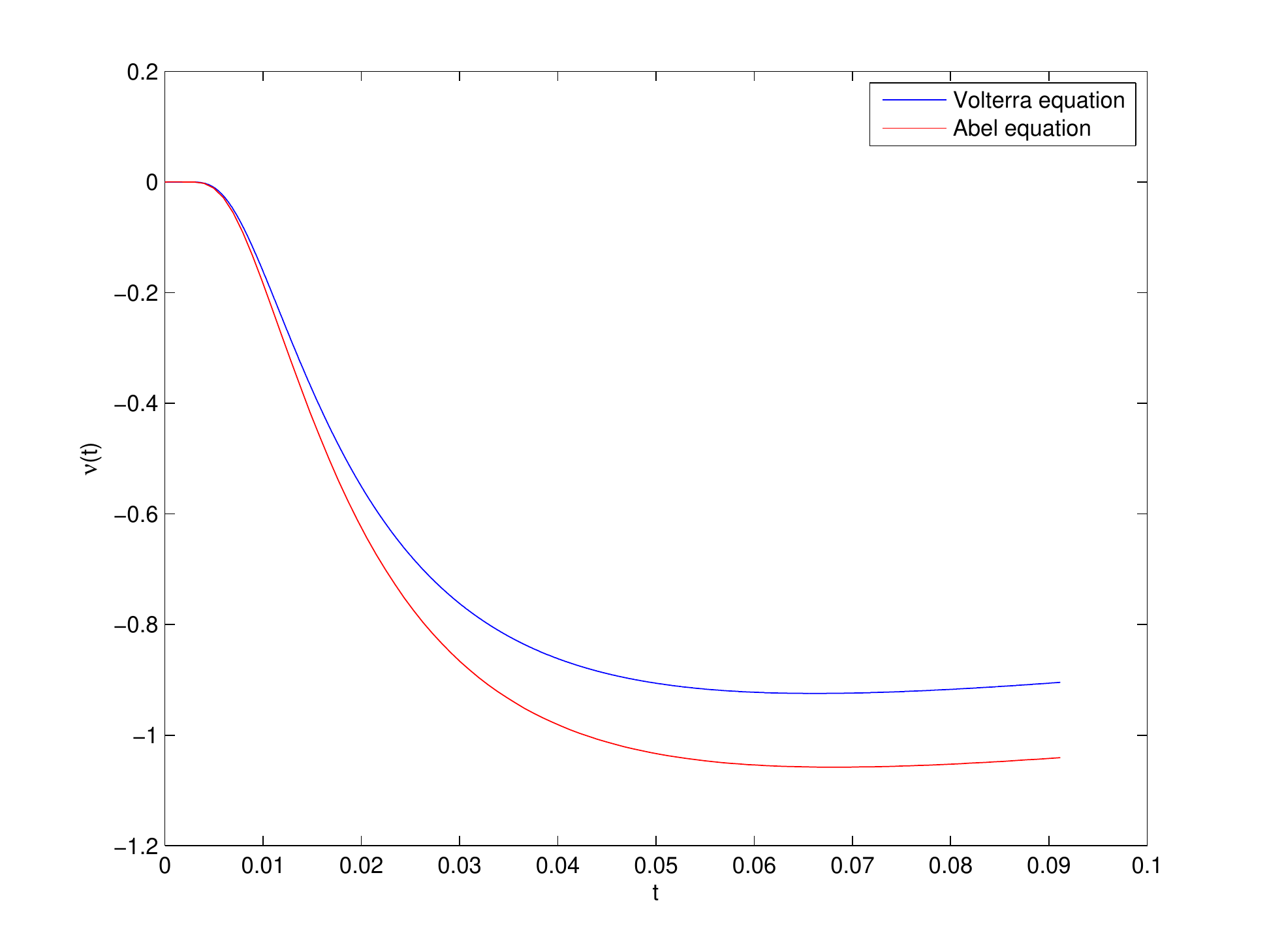}} %
\subfloat[]{\includegraphics[width=0.5\textwidth]{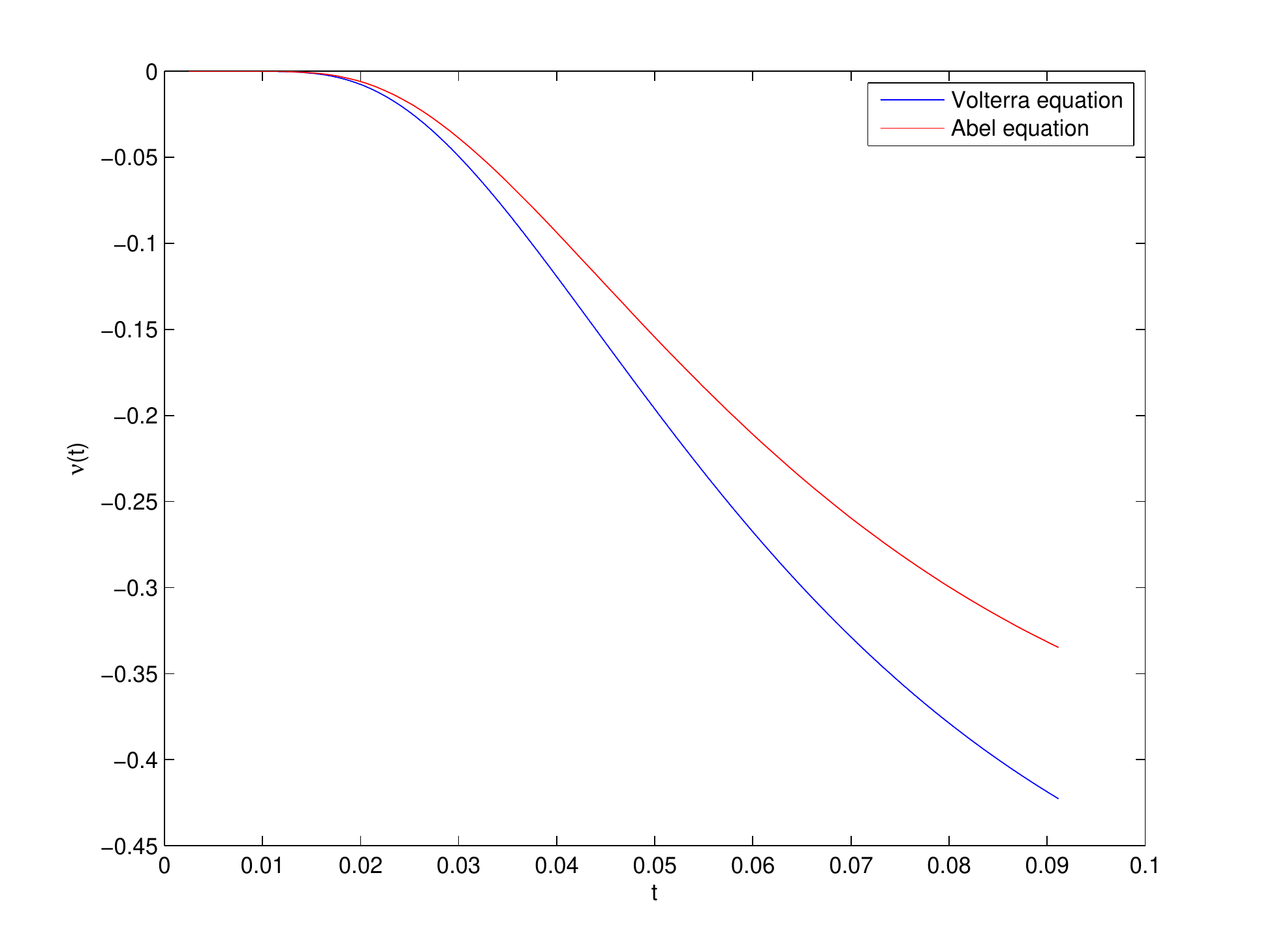}}
\end{center}
\par
\vspace{-10pt}
\caption{ Comparison between the approximation by the Abel equation and the
numerical solution of the forward equation $\protect\nu^f(t)$ in $t$%
-coordinates (a) $b = -0.5, z = -0.25$ (b) $b = 0.5, z = 1$.}
\label{abel}
\end{figure}

\begin{figure}[]
\begin{center}
\subfloat[]{\includegraphics[width=0.5\textwidth]{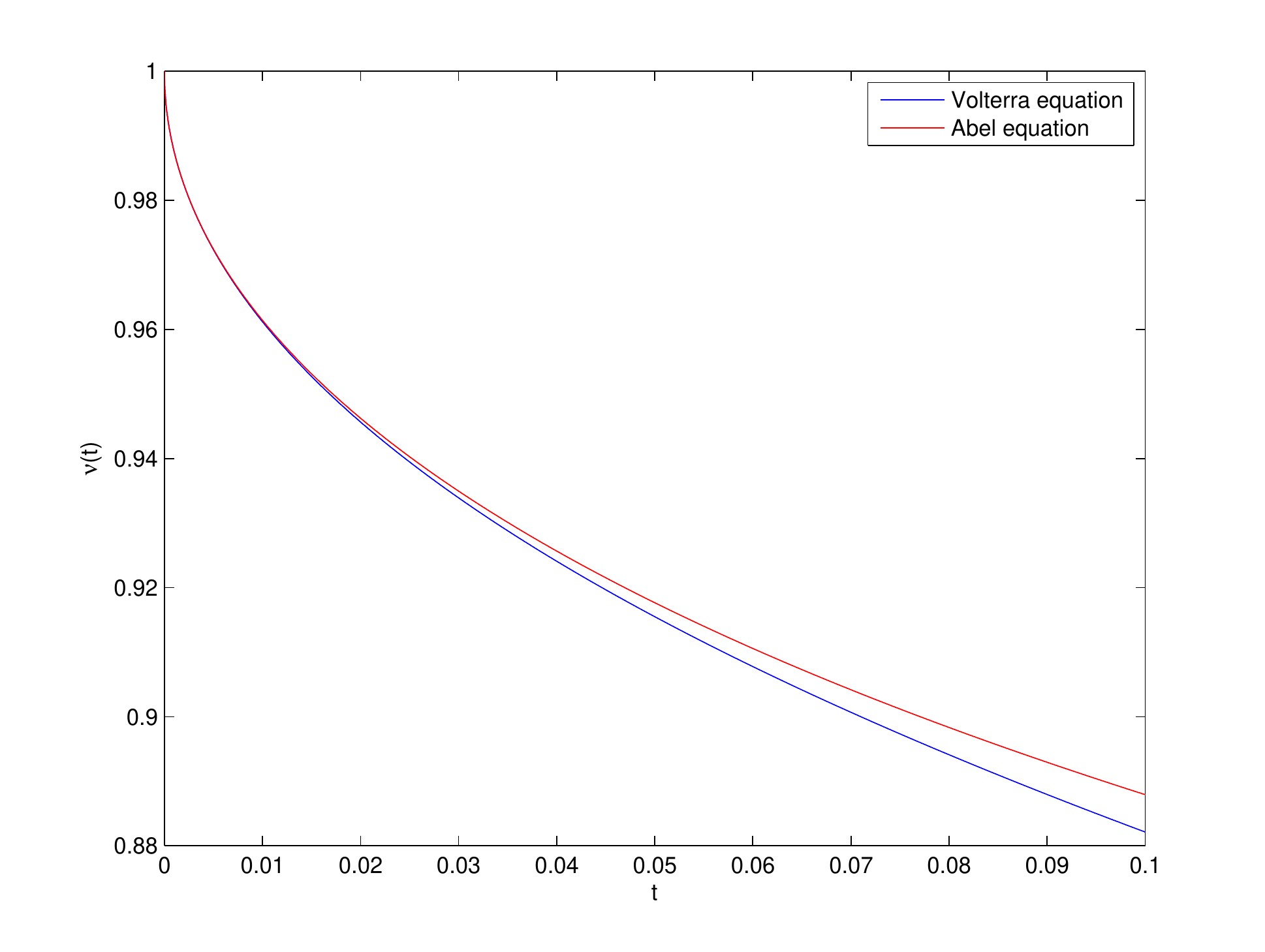}} %
\subfloat[]{\includegraphics[width=0.5\textwidth]{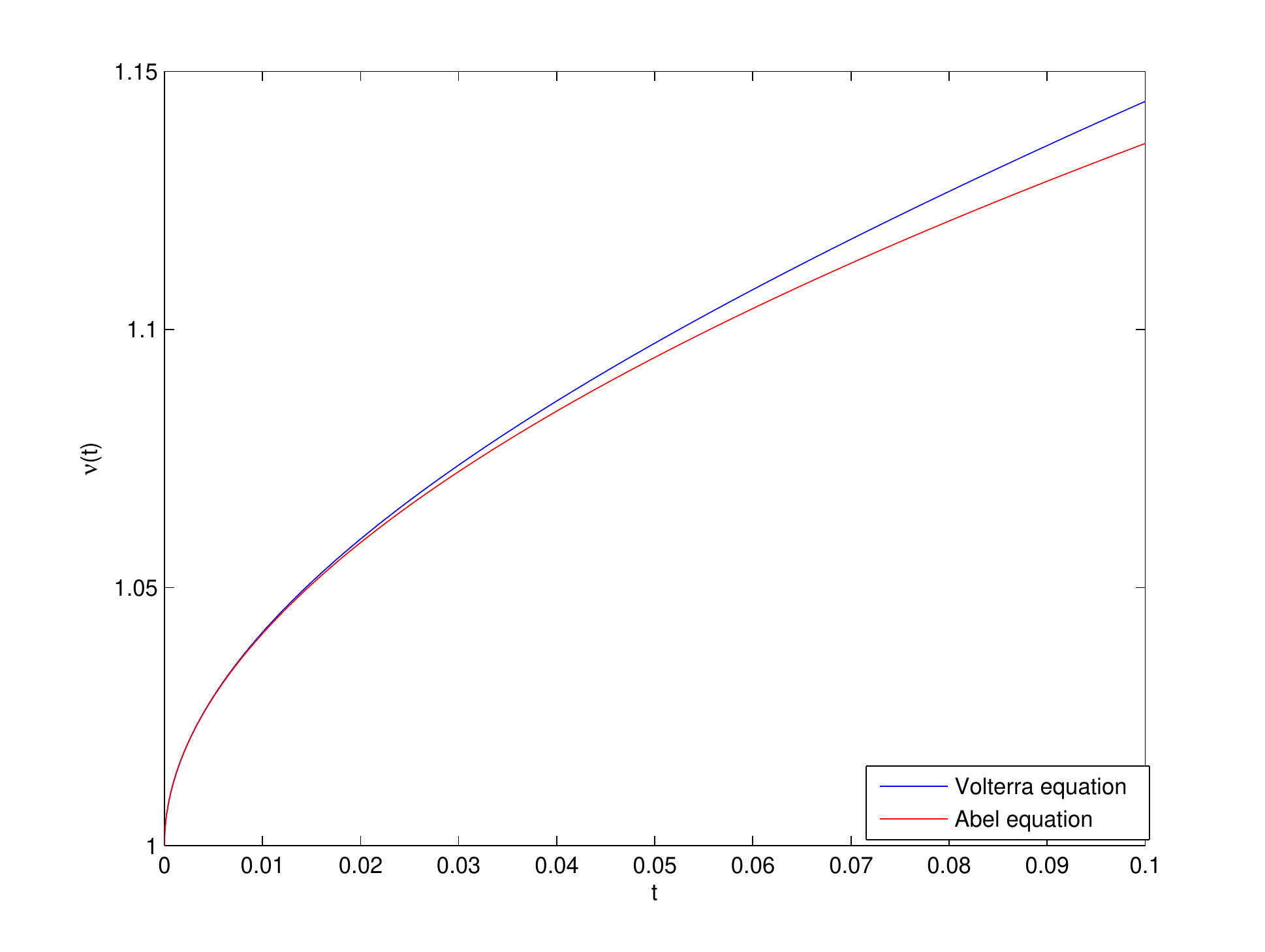}}
\end{center}
\par
\vspace{-10pt}
\caption{ Comparison between the approximation by the Abel equation and the
numerical solution of the backward equation $\protect\nu^b(t)$ in $t$%
-coordinates (a) $b = -0.5$ (b) $b = 0.5$.}
\label{abel_b}
\end{figure}

To test how good the Abel equation approximates the corresponding Volterra equation, we plot them together for small values of $t$. In Figure \ref{abel} we compare the corresponding forward equations and in Figure \ref{abel_b} we compare the corresponding backward equations. We can see that up to $t = 0.02$ the solutions are visually indistinguishable.

\begin{figure}[]
\begin{center}
\subfloat[]{\includegraphics[width=0.5\textwidth]{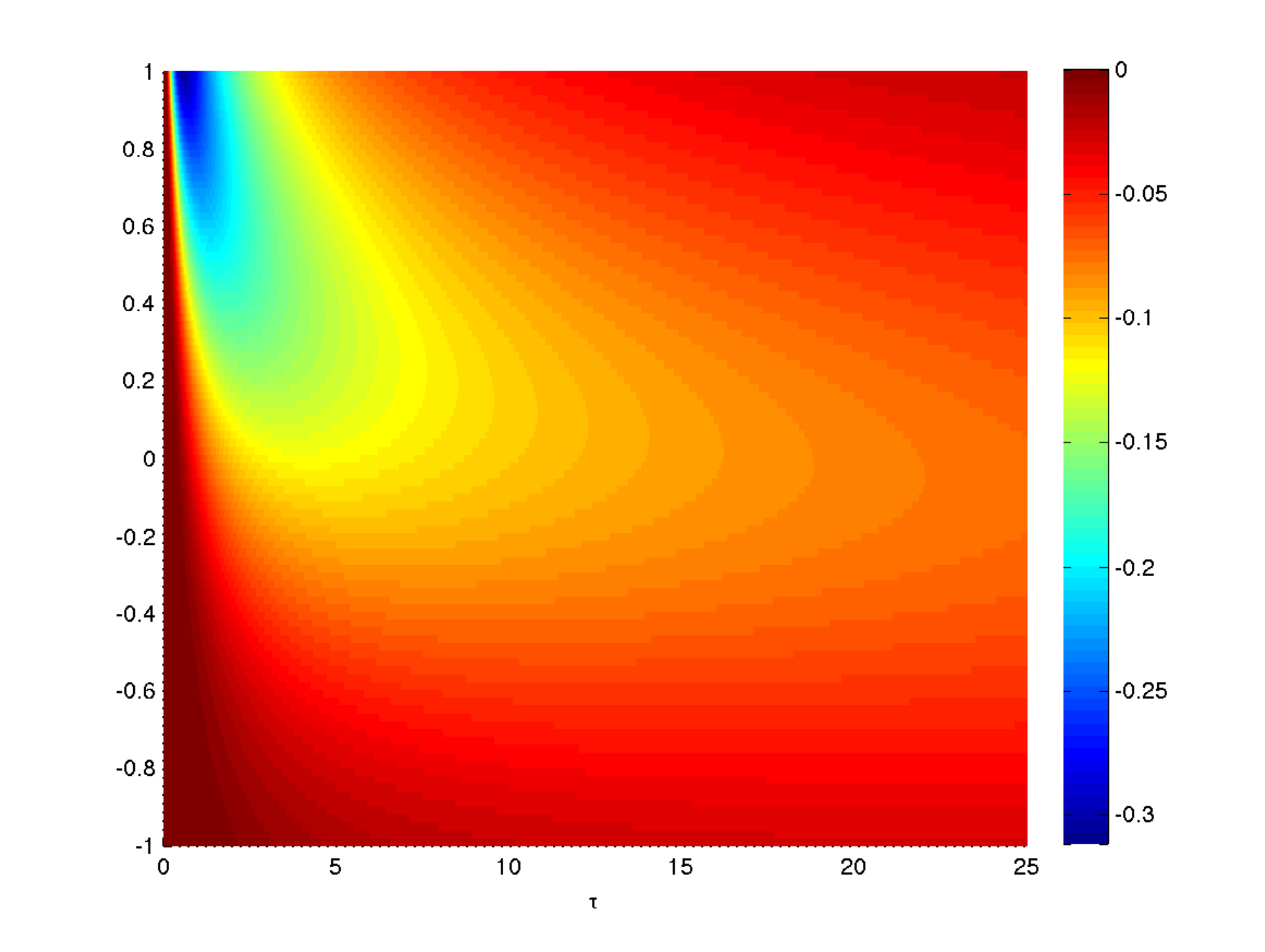}} %
\subfloat[]{\includegraphics[width=0.5\textwidth]{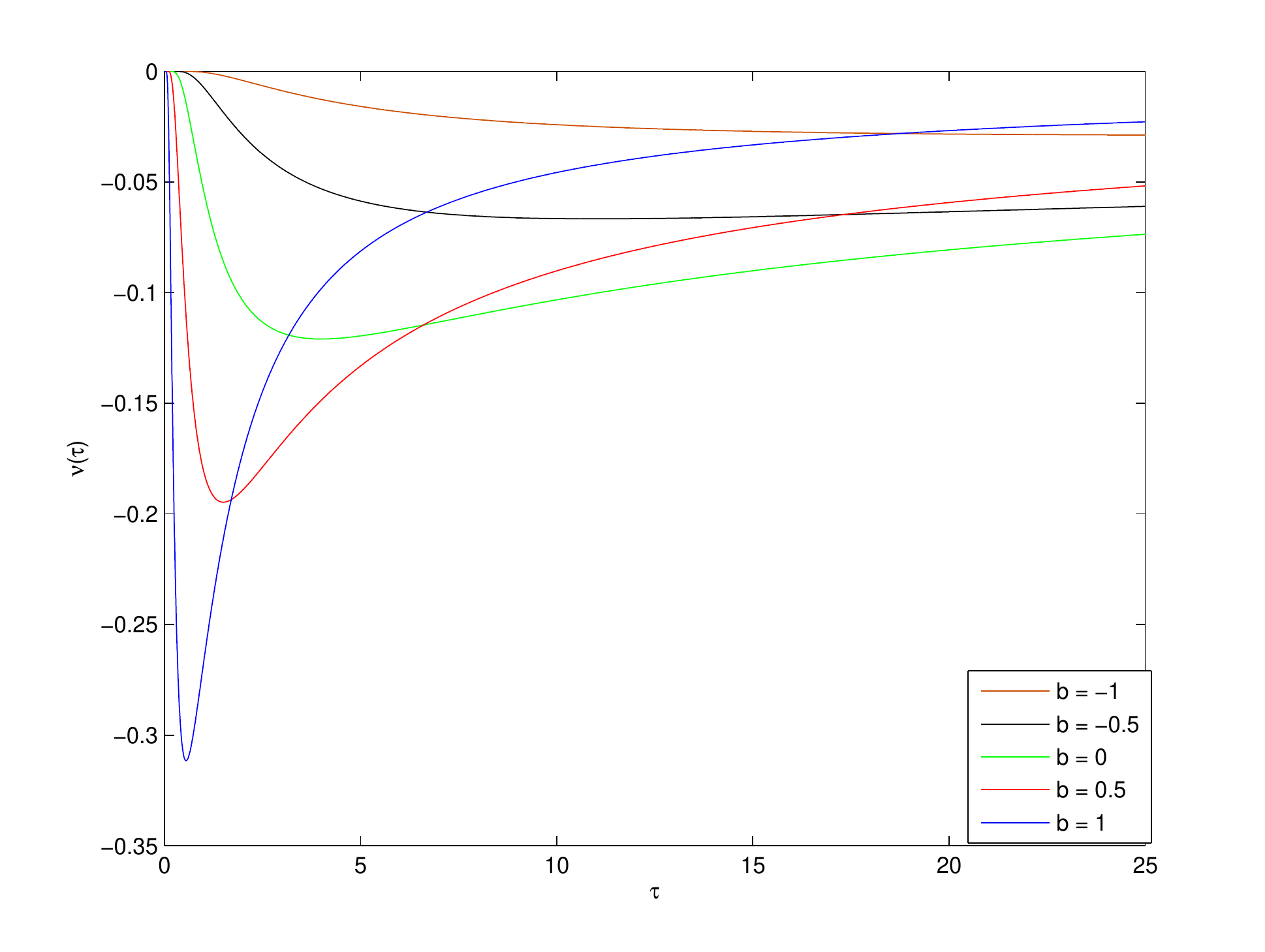}}
\end{center}
\par
\vspace{-10pt}
\caption{ $\protect\nu^f(\protect\theta)$, the solution of \eqref{Eq23a}
with $z = 2$ (a) as a function of $\protect\theta$ and $b$ (b) as a function
of $\protect\theta$ for different values of $b$.}
\label{nu_forward}
\end{figure}

\begin{figure}[]
\begin{center}
\subfloat[]{\includegraphics[width=0.5\textwidth]{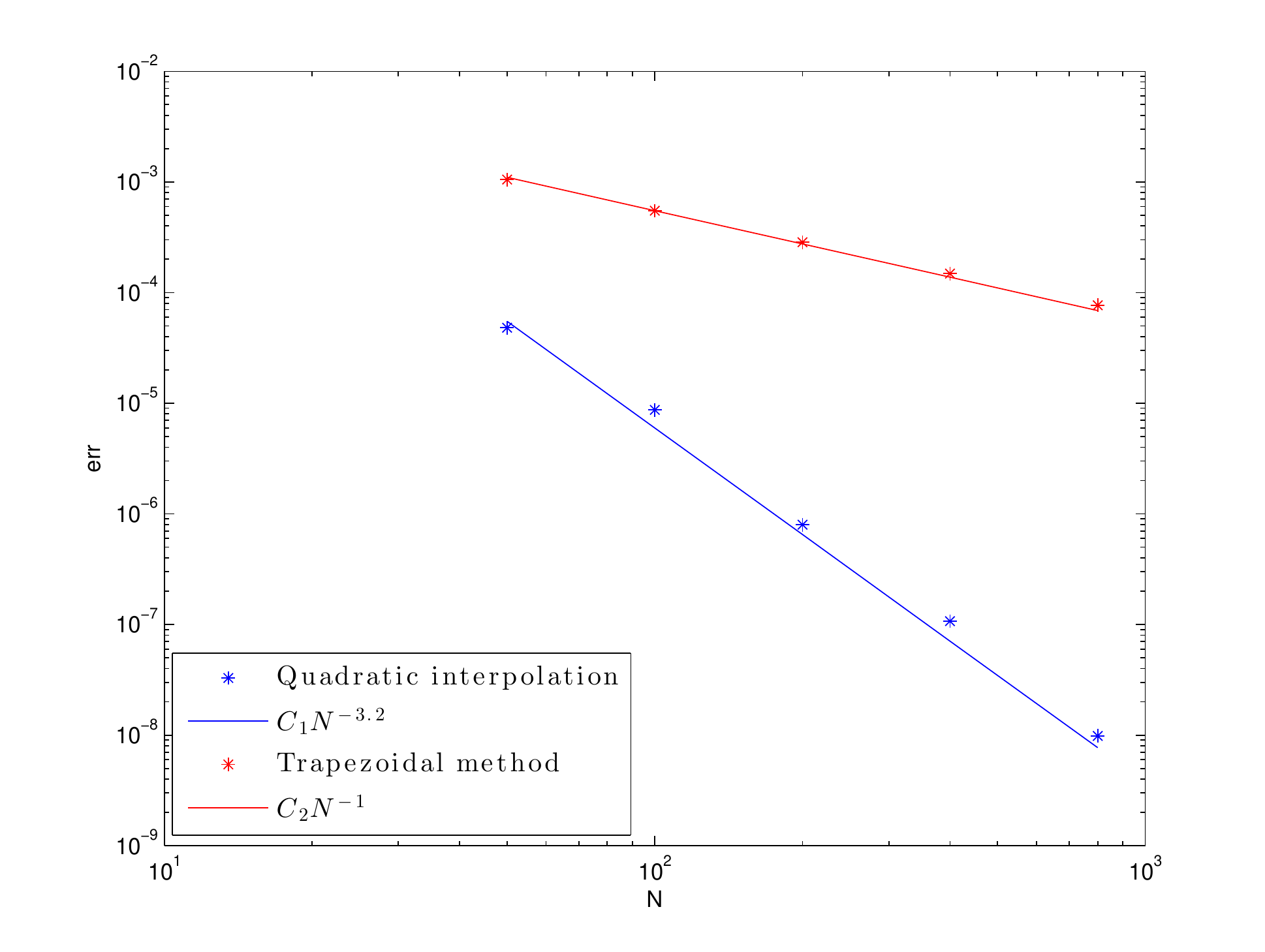}} %
\subfloat[]{\includegraphics[width=0.5\textwidth]{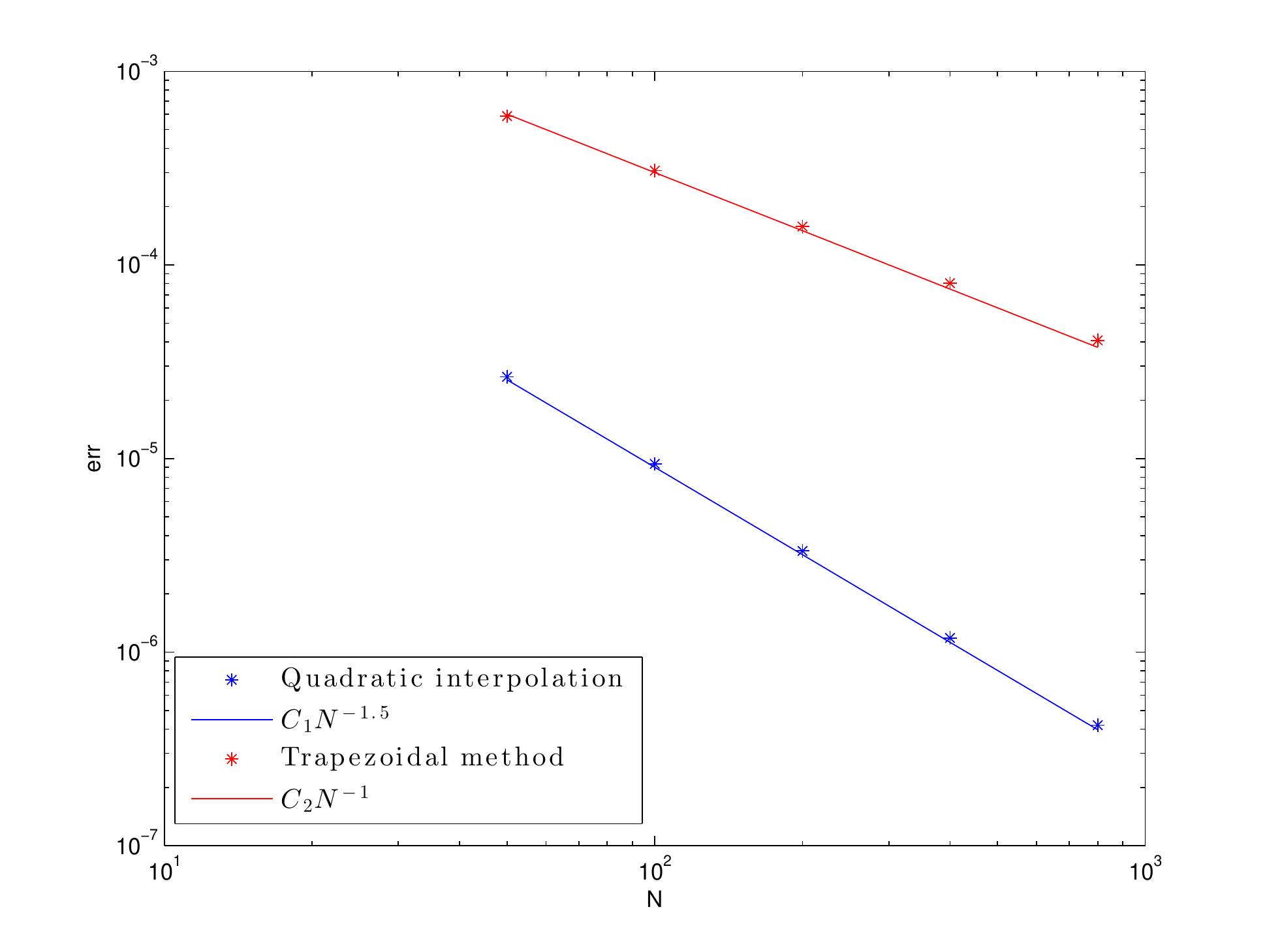}}
\end{center}
\par
\vspace{-10pt}
\caption{ Empirical estimation of the numerical methods from Section \protect
\ref{section:num_method}: (a) forward equation, (b) backward equation.}
\label{conv_rate}
\end{figure}

We also empirically test the convergence rate of our numerical method by
implementing it for the forward and backward Abel equations \eqref{abel_eq_f}
and \eqref{abel_eq} and comparing them with the analytical solutions %
\eqref{abel_analytical_f} and \eqref{abel_analytical}. In Figure \ref%
{conv_rate}a we get the order of convergence 3.2 for the quadratic
interpolation method and order 1 for the trapezoidal method for the forward
equation. It confirms our estimate in Section \ref{section:num_method}. In
Figure \ref{conv_rate}b we get the order 1.5 for the quadratic
interpolation method and the order 1 for the trapezoidal method for the
backward equation. The convergence order of the quadratic interpolation
method for the backward equation is smaller than the theoretical estimate,
because the assumptions on the regularity of the solution are not satisfied.
Potentially, a non-uniform grid improves the convergence order. 

\subsection{Comparison with other methods}

\subsubsection{Description of other methods}

We compare our method with other available alternatives.

\begin{enumerate}
\item \cite{leblanc2000correction} method. We used (3) in \cite%
{leblanc2000correction} to compute the hitting density. In our notation it
is 
\begin{eqnarray*}
g(t) &=&\frac{\left( z-b\right) \exp \left( \frac{z^{2}-b^{2}+t-(z-b)^{2}%
\coth {t}}{2}\right) }{\sqrt{2\pi \left( \sinh \left( t\right) \right) ^{3}}}
\\
&=&\frac{\left( z-b\right) \exp \left( b(z-b)-\frac{e^{-t}\left( z-b\right)
^{2}}{2\sinh \left( t\right) }+\frac{t}{2}\right) }{\sqrt{2\pi \left( \sinh
\left( t\right) \right) ^{3}}}.  \notag
\end{eqnarray*}%
Integrating $g(t)$, we get 
\begin{equation*}
G(t)=2e^{b(z-b)}N\left( -\frac{(z-b)e^{-t/2}}{\sqrt{\sinh {t}}}\right).
\end{equation*}

\item \cite{alili2005representations},  \cite{linetsky2004computing}, and \cite{ricciardi1988first} method. The method is based on the
inversion of the Laplace transform of the hitting density. In our notation,
the Laplace transform $u(\Lambda)$ has the form 
\begin{equation}
u(\Lambda )=e^{\frac{z^{2}-b^{2}}{2}}\frac{D_{-\Lambda }(z\sqrt{2})}{%
D_{-\Lambda }(b\sqrt{2})},  \label{alili_laplace}
\end{equation}%
where $D_{\nu }(x)$ is the parabolic cylinder function.

 \cite{alili2005representations} found a representation of the inverse Laplace transform as a
series of parabolic cylinder function and its derivatives
\begin{equation*}
	g(t) = e^{(z^2-b^2)/2} \sum_{j = 1}^{\infty} \frac{D_{\nu_{j, b \sqrt{2}}}(z \sqrt{2})}{D'_{\nu_{j, b \sqrt{2}}}(b \sqrt{2})} \exp \left(\nu_{j, b \sqrt{2}} \right),
\end{equation*}
where $D'_{\nu}(x) = \frac{\partial D}{\partial \nu}(x)$ and $\nu_{j, b}$ the ordered sequence of positive zeros of $\nu \rightarrow D_{\nu}(x)$.

However, we used Gaver-Stehfest algorithm (\cite{abate2000introduction}) for the inversion, as
it gives more stable and robust results. Moreover, the representation from \cite{alili2005representations}
works only for $t > t_0$ for some $t_0$, while Gaver-Stehfest algorithm works for all positive $t$. 

Using this method, the density can be expressed as 
\begin{equation*}
g(t)\approx \frac{\ln 2}{t}\sum_{k=1}^{2m}\omega _{k}u\left( \frac{k\ln 2}{t}%
\right) ,
\end{equation*}%
where 
\begin{equation*}
\omega _{k}=(-1)^{m+k}\sum_{j=\lfloor (k+1)/2\rfloor }^{\min (k,m)}\frac{%
j^{m+1}}{m\!}C_{m}^{j}C_{2j}^{j}C_{k-j}^{j}.
\end{equation*}%
As we can see, the method only requires the values of $u$ on the positive real
semi-axis, and from \eqref{alili_laplace} one can observe that $u(\Lambda )$
is non-singular on $\mathbb{R}_{+}$. As an example, in Figure \ref{laplace_u}%
, we plot $u(\Lambda )$ for $\Lambda >0$ with  $z = 2$ and various values of $b$. 

As an example of using Gaver-Stehfest algorithm in finance, one can refer to \cite{lipton2011filling}, where the authors used it for the calibration of a local volatility surface.
\begin{figure}[]
\begin{center}
\includegraphics[width=0.6\textwidth]{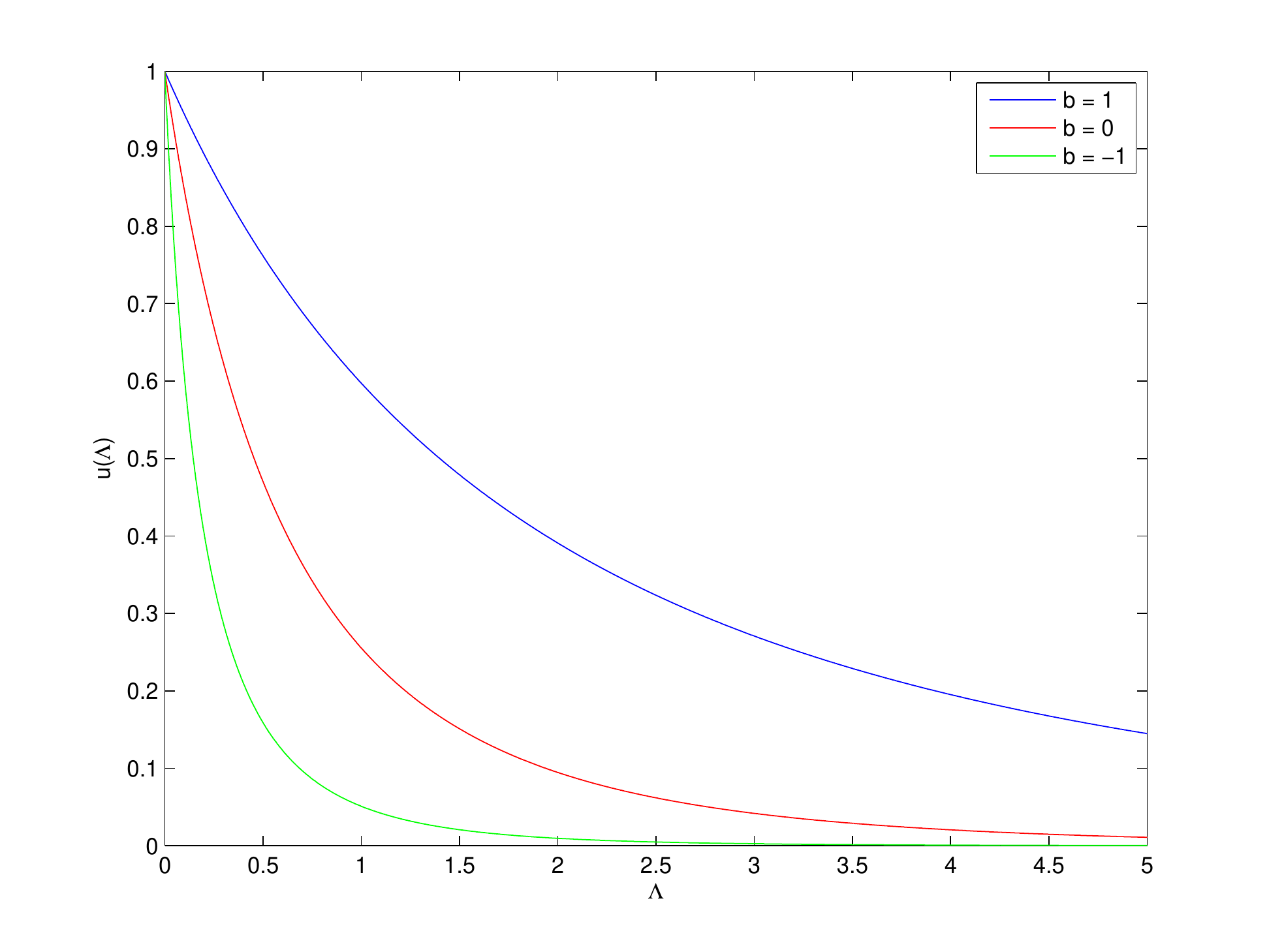}
\end{center}
\par
\vspace{-10pt}
\caption{Laplace transform $u(\Lambda )$ for $z = 2$. }
\label{laplace_u}
\end{figure}

\item Crank--Nicolson method. We solved \eqref{Eq29} using Crank--Nicolson
numerical scheme (\cite{duffy2013finite}).
\end{enumerate}

\subsubsection{Comparison}

\begin{figure}[]
\begin{center}
\subfloat[]{\includegraphics[width=0.5\textwidth]{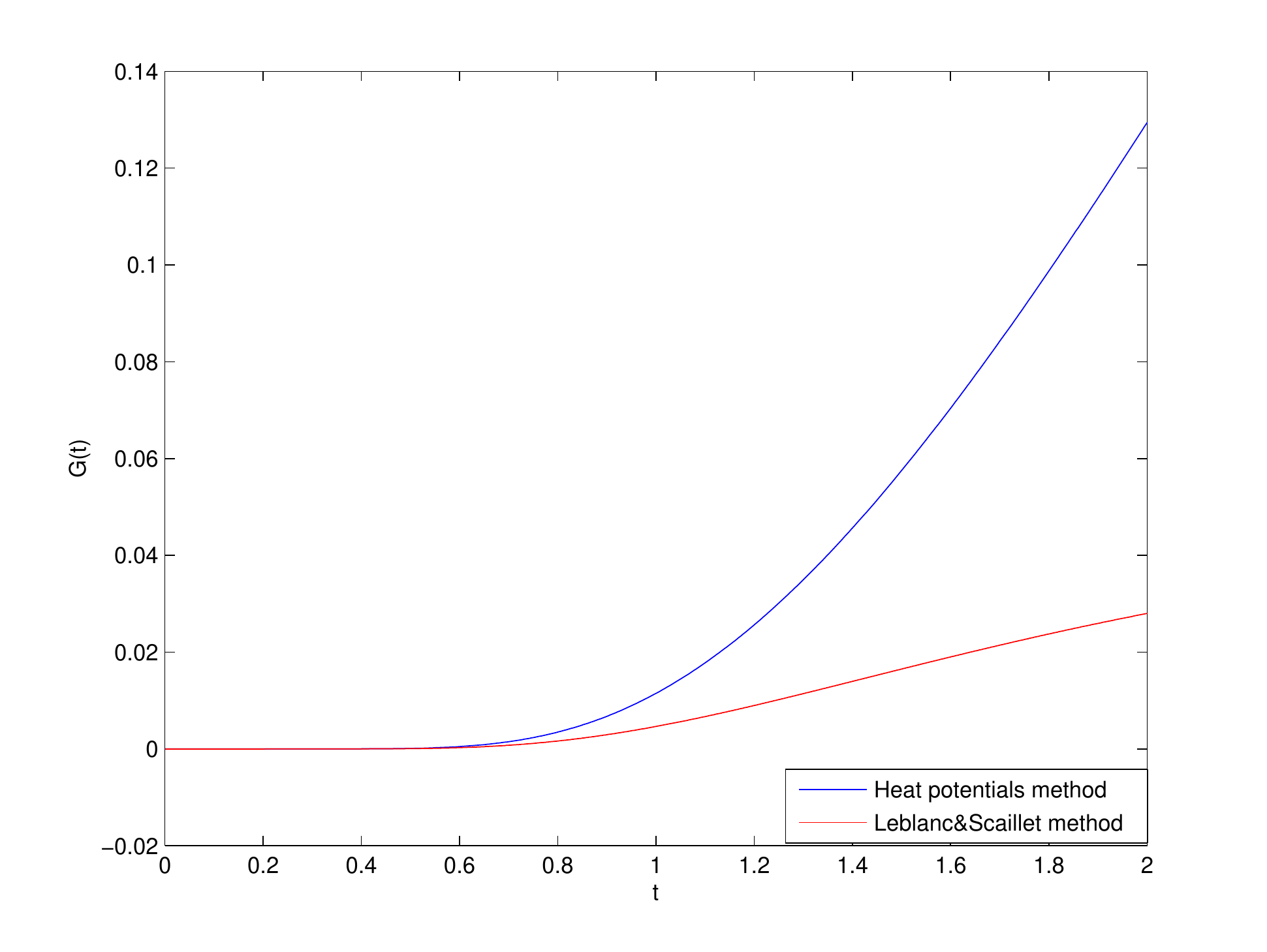}} %
\subfloat[]{\includegraphics[width=0.5\textwidth]{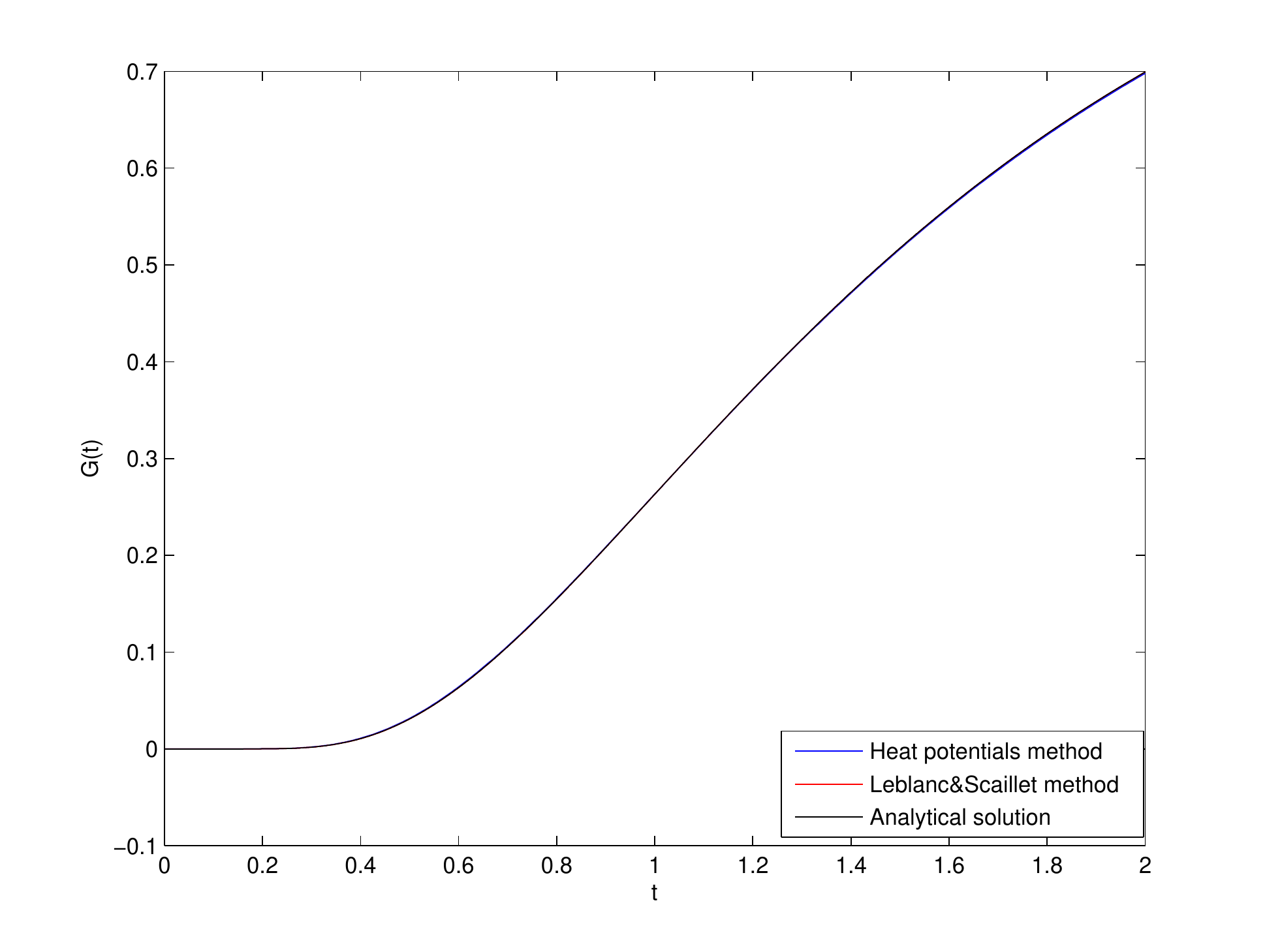}} \\[0pt]
\subfloat[]{\includegraphics[width=0.5\textwidth]{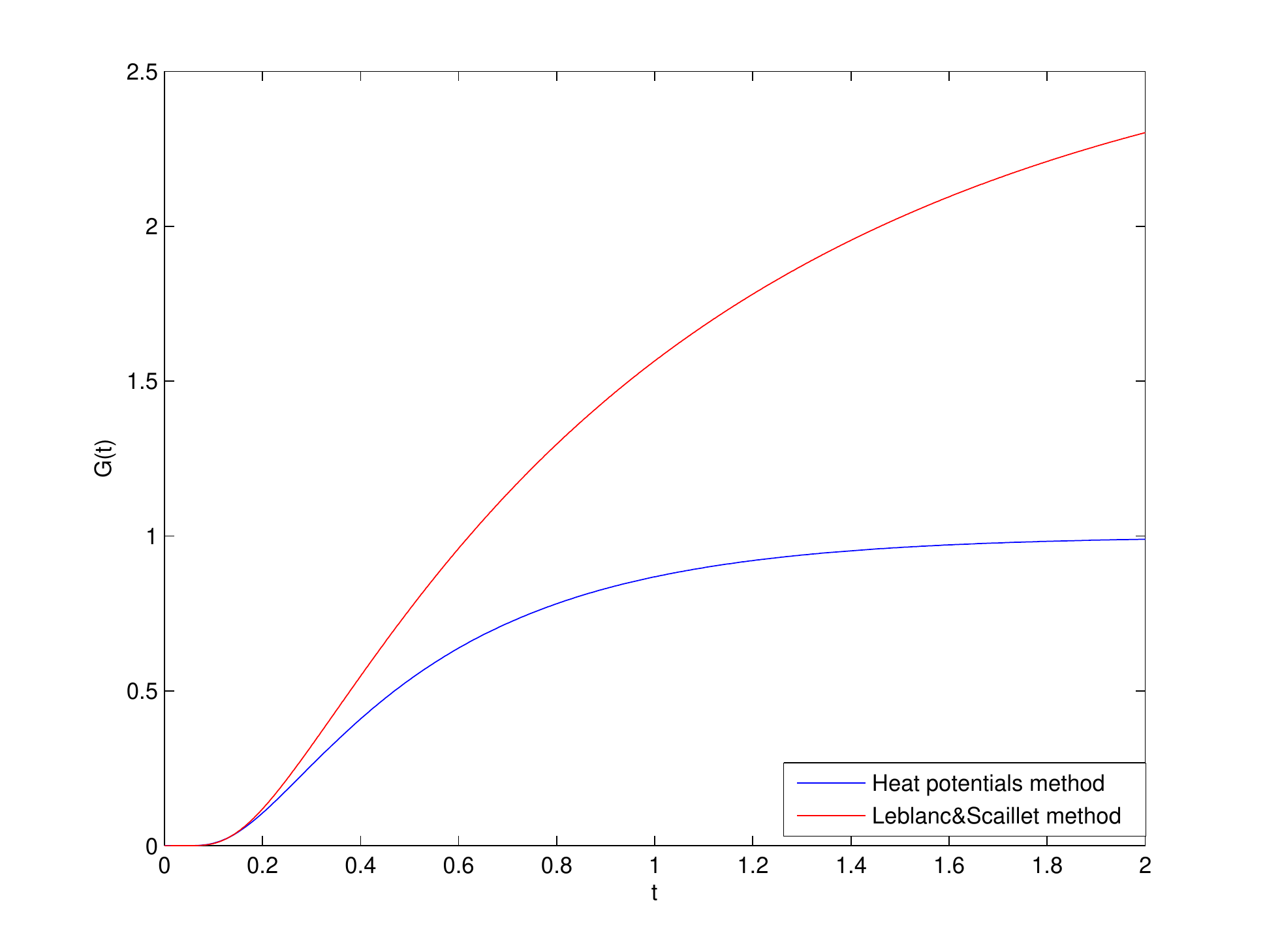}}
\end{center}
\par
\vspace{-10pt}
\caption{Comparison with \protect\cite{leblanc2000correction} method: (a) $z = 2, b = -1$, (b) $z = 2, b = 0$, (c) $z = 2, b = -1$.}
\label{compare_leblanc}
\end{figure}

We start with a comparison with \cite{leblanc2000correction} method to show
that it gives wrong results. In Figure \ref{compare_leblanc} we compare two
methods for different parameters. We also give the analytical solution, when it is available for $b = 0$. We can observe that only for $b = 0$ the \cite%
{leblanc2000correction} method gives correct results, while for other
parameters it totally differs from our method. Moreover in Figure \ref%
{compare_leblanc}c we can see that it gives $G(t) > 1$.

\begin{figure}[]
\begin{center}
\subfloat[]{\includegraphics[width=0.5\textwidth]{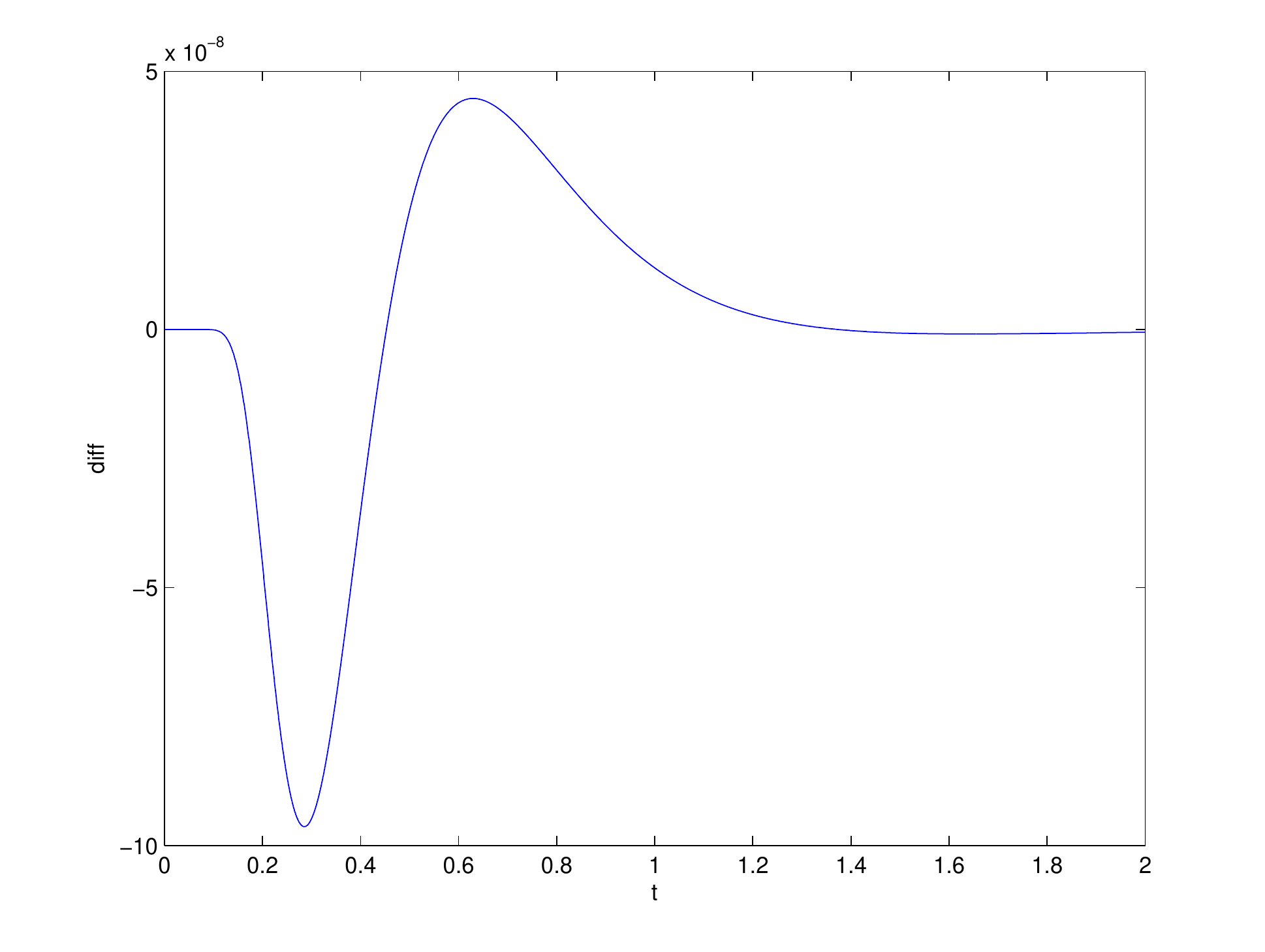}} %
\subfloat[]{\includegraphics[width=0.5\textwidth]{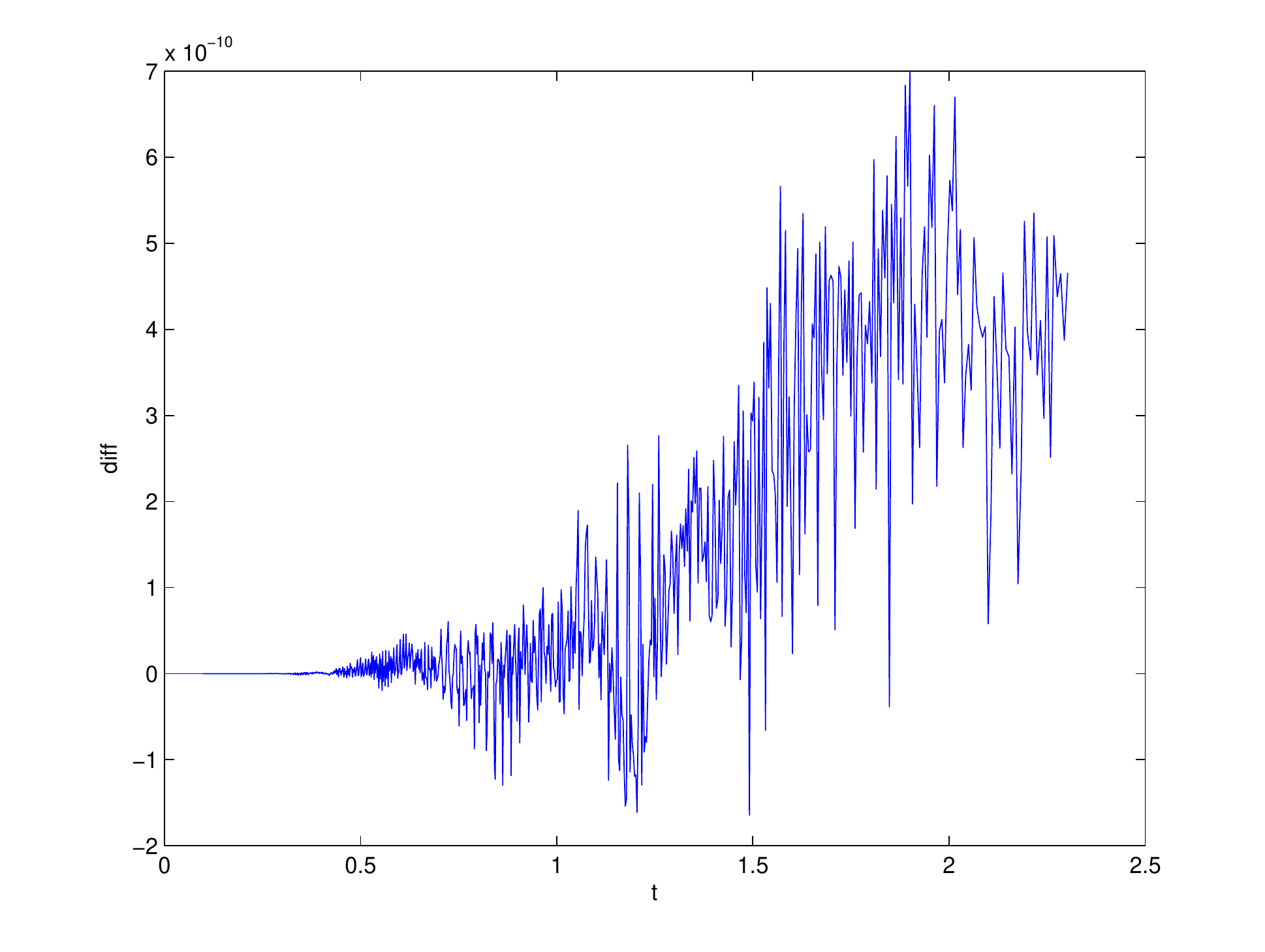}}
\end{center}
\par
\vspace{-10pt}
\caption{ The difference between numerical and analytical solutions for $b =
0$ (a) Forward method (b) Backward method.}
\label{fig_diff}
\end{figure}

\begin{figure}[]
\begin{center}
\subfloat[]{\includegraphics[width=0.5\textwidth]{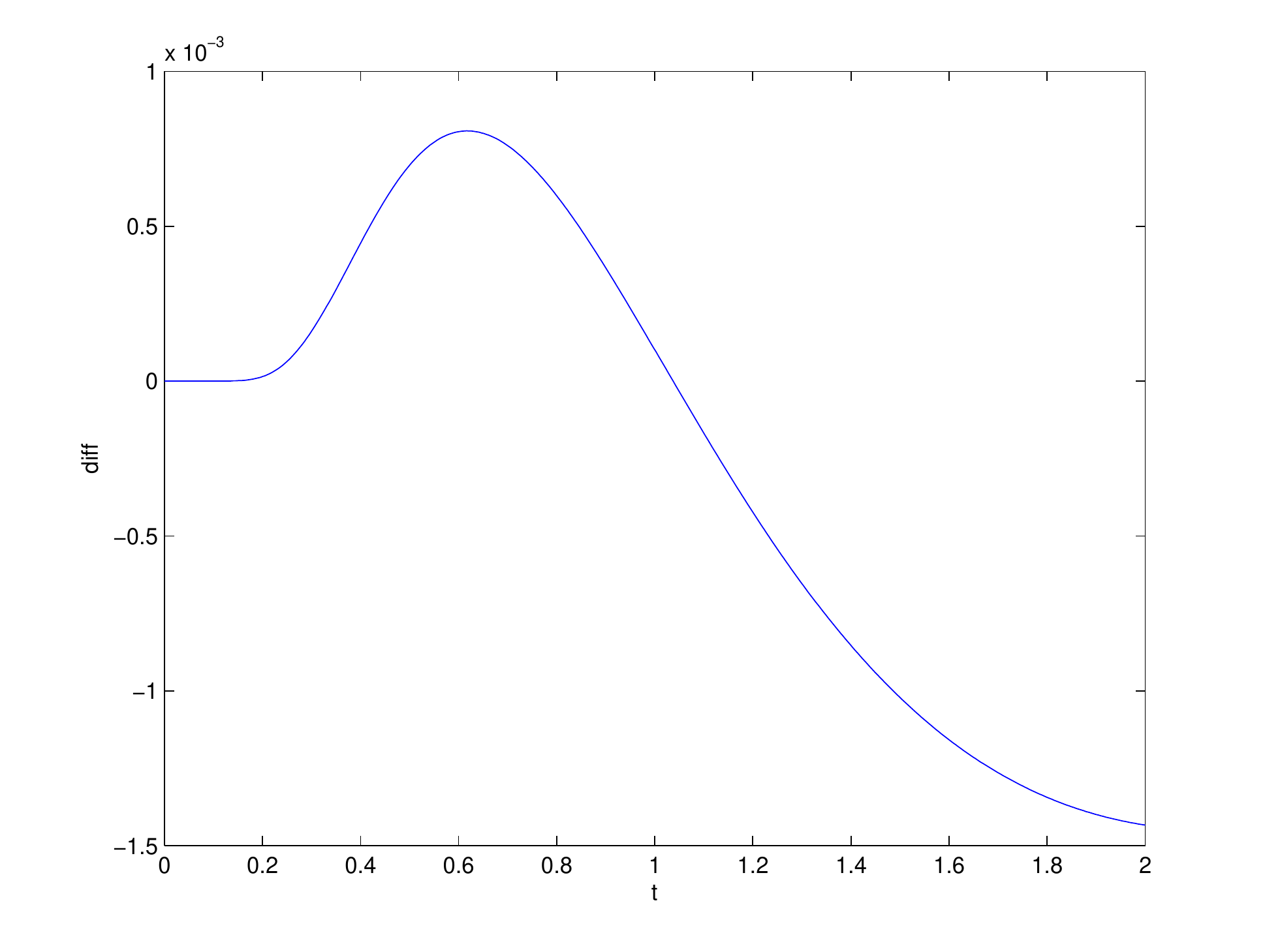}} %
\subfloat[]{\includegraphics[width=0.5\textwidth]{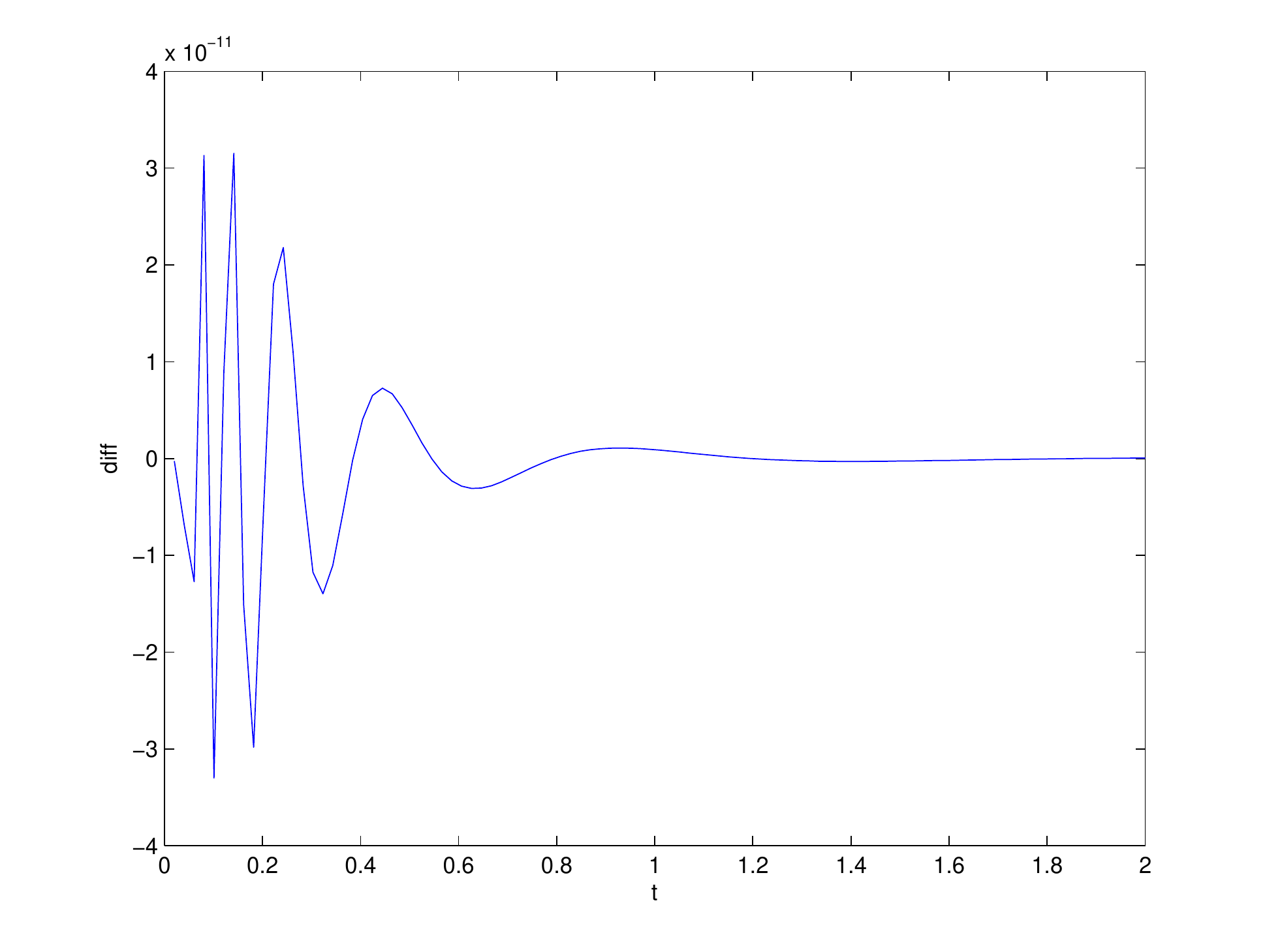}}
\end{center}
\par
\vspace{-10pt}
\caption{ The difference between numerical and analytical solutions (a)
Crank-Nicolson method (b) Laplace transform method.}
\label{fig_diff2}
\end{figure}

A closed-form solution is only available when $b = 0$ and is given by %
\eqref{Eq12}. We take other parameters as before $T = 2$ and $z = 2$. We
compare our method (forward and backward), the Crank-Nicolson method, and the
method based on the inversion of the Laplace transform in Figure \ref{fig_diff} and
Figure \ref{fig_diff2}. We note that our method has an advantage for this
case, because Volterra equations \eqref{Eq23} and \eqref{Eq38a} become
trivial for $b = 0$. Nevertheless, the comparison is still very useful.

We take $N = 500$ for both forward and backward methods, $h = k = 0.005$
for the Crank-Nicolson method, and $m = 24$ for the Gaver-Stehfest algorithm.
 We used  \verb mpmath  library in \verb Python  (\cite{mpmath})  for the implementation of the Gaver-Stehfest algorithm and
  parabolic cylinder functions with arbitrary precision arithmetics. In our computations we used $100$ digits precision.
  
 From these graphs we observe that the methods
developed in this paper and the method based on the inversion of the Laplace transform (\cite{alili2005representations})
significantly outperform the Crank-Nicolson scheme. 

\begin{figure}[]
\begin{center}
\subfloat[]{\includegraphics[width=0.5\textwidth]{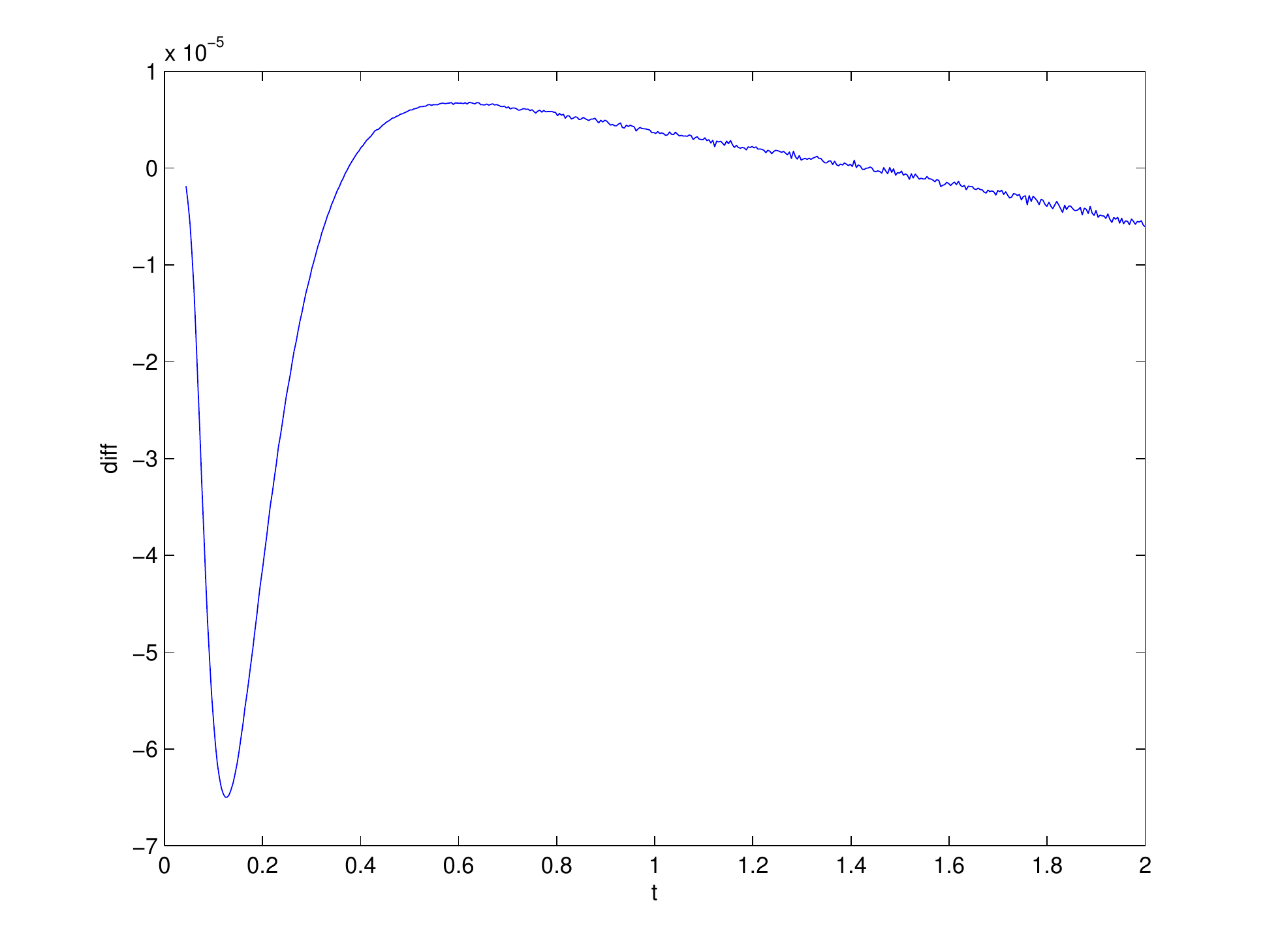}} %
\subfloat[]{\includegraphics[width=0.5\textwidth]{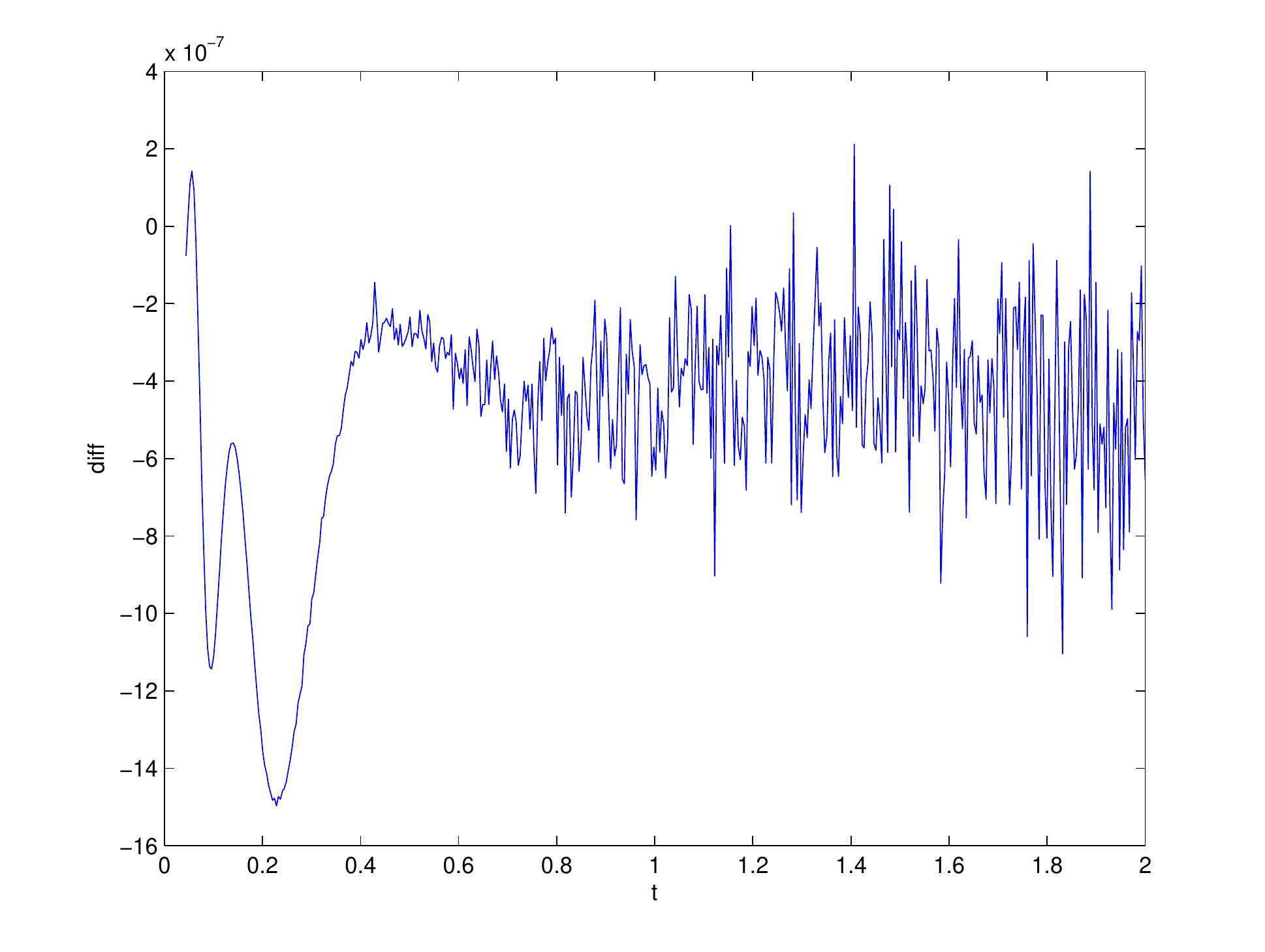}}
\end{center}
\par
\vspace{-10pt}
\caption{ The difference between our method and the Laplace transform method
(a) $N = 100$ (b) $N = 10000$.}
\label{fig_diff3}
\end{figure}

In order to compare the solutions for non-trivial parameters, we take the
Laplace transform method with a maximum possible precision, and our method
for $N = 100$ and $N = 10000$. The first example shows how our results can
be comparable to the Laplace transform method for a relatively small value
of $N$, when the computations can be done very fast; the second example
demonstrates that our method can obtain a very good precision for a large $N$%
. The results are presented in Figure \ref{fig_diff3}. 

\subsection{Asymptotic behavior when $t \to \infty$}
\begin{figure}[]
\begin{center}
\subfloat[]{\includegraphics[width=0.5\textwidth]{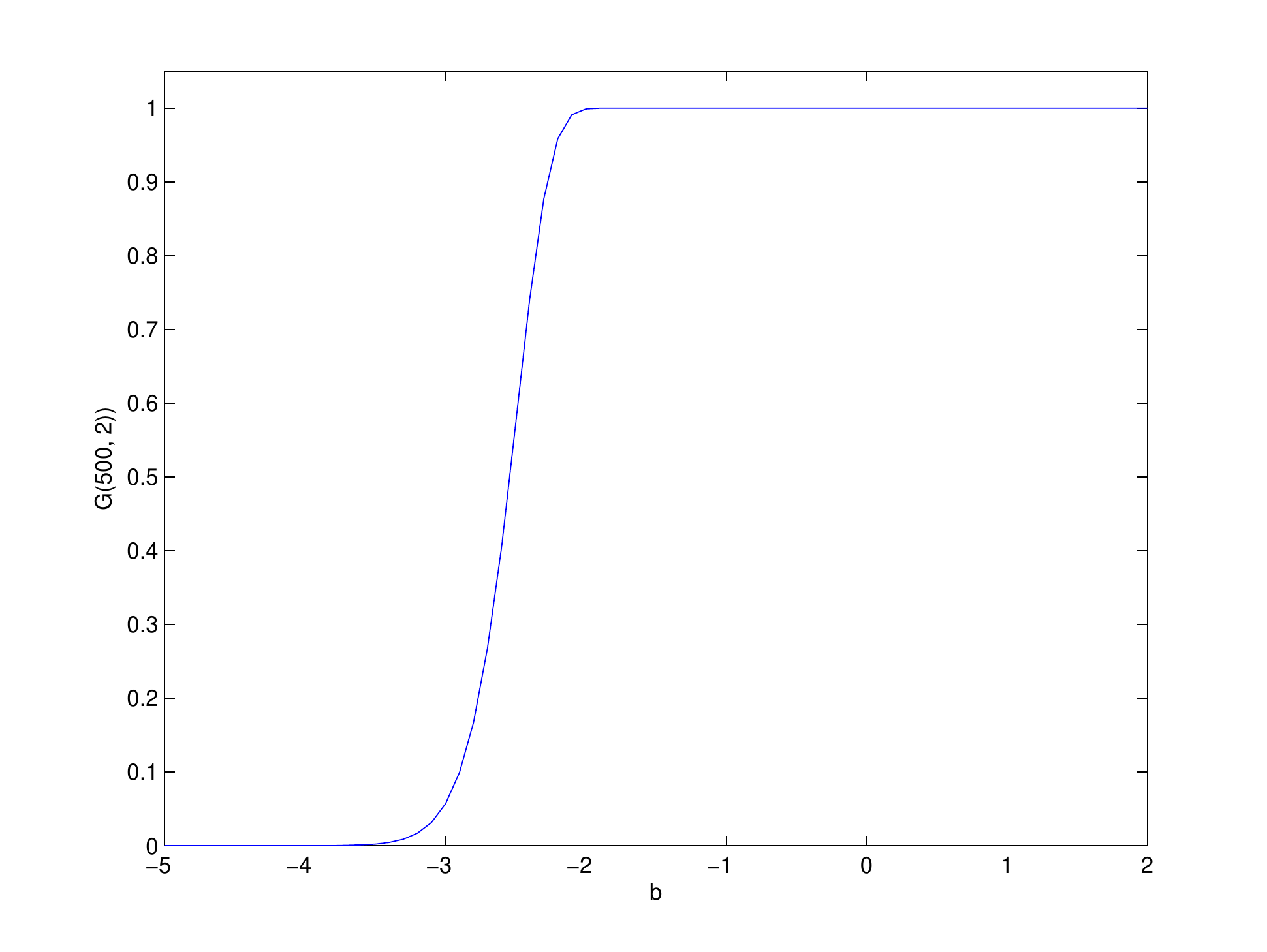}\label{fig_asympt}}
\subfloat[]{\includegraphics[width=0.5\textwidth]{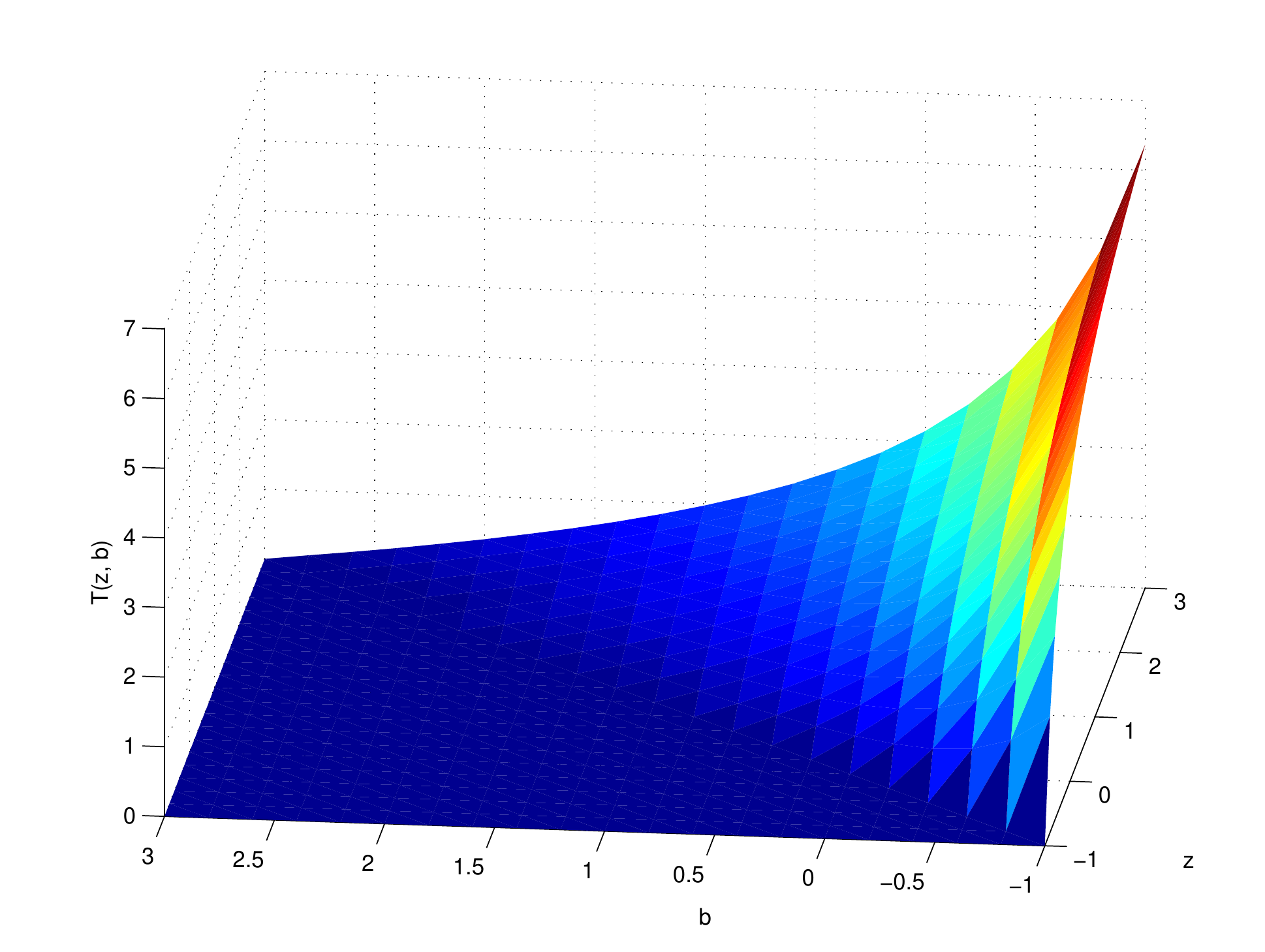} \label{T_z_b}}

\end{center}
\par
\vspace{-10pt}
\caption{(a) $G(T, z)$ as a function of $b$ with $T = 500$ and $z = 2$ (b) $T(z, b) = \mathbb{E}[s]$, where $s$ is the hitting time.}
\end{figure}

It is clear that for $b= 0$, and hence all $b \ge 0$, $G(t, z) \to 1$, when $t \to \infty$. One can easily verify it by taking the limit in \eqref{Eq40}. But it is unclear what happens for a negative $b$. We empirically investigate it by plotting $G(t, z)$ as a function of $b$ for a fixed large value of $T$ and fixed $z$. In Figure \ref{fig_asympt} we take $z = 2$ and $T = 500$ and compute $G(T, z)$ for $b \in [-5, 2]$. We can see that it still remains close to $1$ up to $b = -2$, and then rapidly approaches zero. In further research we want to explore the asymptotic behavior of $G(t)$ in more details.  In Figure \ref{T_z_b} we show the expected hitting time; this quantity is very important for the design of mean-reverting trading strategies since it allows a trader to decide when to go in and out of the trade.

\section{Conclusion}

\label{sec:conclusion} We developed a semi-analytical approach to finding
the first time hitting density of an Ornstein-Uhlenbeck process. We
transformed our problem to a free-boundary value problem; using the method
of heat potentials, we derived the corresponding expression for the density
via a weight function, which is a solution of a Volterra equation of the
second kind. For small values of $t$ we solved this equation analytically by
using Abel equation approximation, while for $t$ which are not small, we
developed a numerical recursive scheme. We showed the third order of
convergence for the forward scheme and the order 1.5 for the backward scheme and confirmed it numerically.

We compared our method to several other methods known in the literature. We
showed that our method significantly outperforms the Crank-Nicolson scheme and
is at least as good as the Laplace transform method and is markedly more stable.

As we pointed out, the problem has many practical applications, 
especially for construction of mean-reverting trading strategies. 
Using the surface Figure \ref{T_z_b}, one can estimate the expected time
 for the reversion to the mean, and construct either a long or a short strategy.

This paper leaves several important questions open as to the asymptotic behavior of the hitting time density, which we are going to address in further research.
\newpage

\appendix

\section{Derivation of \eqref{Eq22}}

\label{appendix} 
\begin{eqnarray*}
g\left( t\right) &=&\frac{1}{2}p_{x}\left( t,b\right) =\frac{1}{2}%
\lim_{x\rightarrow b}p_{x}(t,x)= \\
&=&-\frac{\left( e^{t}b-z\right) \exp \left( -\frac{\left( e^{t}b-z\right)
^{2}}{\left( e^{2t}-1\right) }+2t\right) }{\sqrt{\pi \left( e^{2t}-1\right)
^{3}}} \\
&&+\frac{1}{2}e^{2t}\lim_{\varepsilon \rightarrow 0}\frac{\partial }{%
\partial \varepsilon }\int_{0}^{\tau }\frac{\left( \varepsilon +b\left( \tau
\right) -b\left( \tau ^{\prime }\right) \right) \exp \left( -\frac{\left(
\varepsilon +b\left( \tau \right) -b\left( \tau ^{\prime }\right) \right)
^{2}}{2\left( \tau -\tau ^{\prime }\right) }\right) }{\sqrt{2\pi \left( \tau
-\tau ^{\prime }\right) ^{3}}}\nu \left( \tau ^{\prime }\right) d\tau
^{\prime } \\
&=&-\frac{\left( e^{t}b-z\right) \exp \left( -\frac{\left( e^{t}b-z\right)
^{2}}{\left( e^{2t}-1\right) }+2t\right) }{\sqrt{\pi \left( e^{2t}-1\right)
^{3}}} \\
&&+\frac{1}{2}e^{2t}\lim_{\varepsilon \rightarrow 0}\int_{0}^{\tau }\left( 1-%
\frac{\left( \varepsilon +b\left( \tau \right) -b\left( \tau ^{\prime
}\right) \right) ^{2}}{\left( \tau -\tau ^{\prime }\right) }\right) \frac{%
\exp \left( -\frac{\left( \varepsilon +b\left( \tau \right) -b\left( \tau
^{\prime }\right) \right) ^{2}}{2\left( \tau -\tau ^{\prime }\right) }%
\right) }{\sqrt{2\pi \left( \tau -\tau ^{\prime }\right) ^{3}}}\nu \left(
\tau ^{\prime }\right) d\tau ^{\prime }. \\
&&
\end{eqnarray*}%
We compute the limit in the second term separately%
\begin{eqnarray*}
&&\lim_{\varepsilon \rightarrow 0}\int_{0}^{\tau }\left( 1-\frac{\left(
\varepsilon +b\left( \tau \right) -b\left( \tau ^{\prime }\right) \right)
^{2}}{\left( \tau -\tau ^{\prime }\right) }\right) \frac{\exp \left( -\frac{%
\left( \varepsilon +b\left( \tau \right) -b\left( \tau ^{\prime }\right)
\right) ^{2}}{2\left( \tau -\tau ^{\prime }\right) }\right) }{\sqrt{2\pi
\left( \tau -\tau ^{\prime }\right) ^{3}}}\nu \left( \tau ^{\prime }\right)
d\tau ^{\prime } \\
&=&\lim_{\varepsilon \rightarrow 0}\int_{0}^{\tau }\left( 1-\frac{%
\varepsilon ^{2}}{\tau -\tau ^{\prime }}-2\frac{\varepsilon (b(\tau )-b(\tau
^{\prime }))}{\tau -\tau ^{\prime }}-\frac{(b(\tau )-b(\tau ^{\prime 2}}{%
\tau -\tau ^{\prime }}\right) \\
&&\times \frac{\exp \left( -\frac{\varepsilon ^{2}}{2(\tau -\tau ^{\prime })}%
+\frac{\varepsilon (b(\tau )-b(\tau ^{\prime }))}{\tau -\tau ^{\prime }}%
\right) \Xi (\tau ,\tau ^{\prime })}{\sqrt{2\pi \left( \tau -\tau ^{\prime
}\right) ^{3}}}\nu \left( \tau ^{\prime }\right) d\tau ^{\prime } \\
&=&\mathbb{L}_{1}+\mathbb{L}_{2}-2\mathbb{L}_{3}-\mathbb{L}_{4},
\end{eqnarray*}
where 
\begin{align*}
\mathbb{L}_{1}& =\nu (\tau )\lim_{\varepsilon \rightarrow 0}\int_{0}^{\tau
}\left( 1-\frac{\varepsilon ^{2}}{\tau -\tau ^{\prime }}\right) \frac{\exp
\left( -\frac{\varepsilon ^{2}}{2(\tau -\tau ^{\prime })}+\frac{\varepsilon
(b(\tau )-b(\tau ^{\prime }))}{\tau -\tau ^{\prime }}\right) }{\sqrt{2\pi
\left( \tau -\tau ^{\prime }\right) ^{3}}}d\tau ^{\prime }, \\
\mathbb{L}_{2}& =\lim_{\varepsilon \rightarrow 0}\int_{0}^{\tau }\left( 1-%
\frac{\varepsilon ^{2}}{\tau -\tau ^{\prime }}\right) \frac{\exp \left( -%
\frac{\varepsilon ^{2}}{2(\tau -\tau ^{\prime })}+\frac{\varepsilon (b(\tau
)-b(\tau ^{\prime }))}{\tau -\tau ^{\prime }}\right) \left( \Xi (\tau ,\tau
^{\prime })\nu (\tau ^{\prime })-\nu (\tau )\right) }{\sqrt{2\pi \left( \tau
-\tau ^{\prime }\right) ^{3}}}d\tau ^{\prime }, \\
\mathbb{L}_{3}& =\lim_{\varepsilon \rightarrow 0}\int_{0}^{\tau }\frac{%
\varepsilon (b(\tau )-b(\tau ^{\prime }))}{\tau -\tau ^{\prime }}\frac{\exp
\left( -\frac{\varepsilon ^{2}}{2(\tau -\tau ^{\prime })}+\frac{\varepsilon
(b(\tau )-b(\tau ^{\prime }))}{\tau -\tau ^{\prime }}\right) \Xi (\tau ,\tau
^{\prime })}{\sqrt{2\pi \left( \tau -\tau ^{\prime }\right) ^{3}}}\nu \left(
\tau ^{\prime }\right) d\tau ^{\prime } \\
\mathbb{L}_{4}& =\lim_{\varepsilon \rightarrow 0}\int_{0}^{\tau }\frac{%
(b(\tau )-b(\tau ^{\prime 2}}{\tau -\tau ^{\prime }}\frac{\exp \left( -\frac{%
\varepsilon ^{2}}{2(\tau -\tau ^{\prime })}+\frac{\varepsilon (b(\tau
)-b(\tau ^{\prime }))}{\tau -\tau ^{\prime }}\right) \Xi (\tau ,\tau
^{\prime })}{\sqrt{2\pi \left( \tau -\tau ^{\prime }\right) ^{3}}}\nu \left(
\tau ^{\prime }\right) d\tau ^{\prime },
\end{align*}%
and 
\begin{equation*}
\Xi (\tau ,\tau ^{\prime })=\exp \left( -\frac{\left( b\left( \tau \right)
-b\left( \tau ^{\prime }\right) \right) ^{2}}{2\left( \tau -\tau ^{\prime
}\right) }.\right)
\end{equation*}%
We have%
\begin{equation*}
\begin{aligned} \mathbb{{L}}_{1}&= \nu(\tau)\underset{\varepsilon
\rightarrow 0}{\lim }\, \int_{0}^{\tau}\left(
1-\frac{\varepsilon^{2}}{\left( \tau-\tau^{\prime }\right) }\right)
\frac{\exp \left(- \frac{\varepsilon^2}{2(\tau - \tau')} + \frac{\varepsilon
(b(\tau) - b(\tau'))}{\tau - \tau'} \right) }{\sqrt{2\pi \left( \tau -\tau
^{\prime }\right) ^{3}}} \, d \tau^{\prime } \\ &=\nu \left( \tau\right)
\underset{\varepsilon\rightarrow 0}{\lim
}\frac{1}{\varepsilon}\int_{0}^{\tau/\varepsilon^{2}}\left(
1-\frac{1}{u}\right) \frac{\exp \left( -\frac{1}{2u}\right) }{\sqrt{2\pi
u^{3}}}\, du \\ &=2\nu \left( \tau\right) \underset{\varepsilon\rightarrow
0}{\lim }\frac{1}{\varepsilon}\int_{\varepsilon/\sqrt{\tau}}^{\infty }\left(
1-v^{2}\right) \frac{\exp \left( -\frac{v^{2}}{2}\right) }{\sqrt{2\pi }} \,
dv \\ &=-2\nu \left( \tau\right) \underset{\varepsilon\rightarrow 0}{\lim
}\frac{1}{\varepsilon}\int_{0}^{\varepsilon/\sqrt{\tau}}\left(
1-v^{2}\right) \frac{\exp \left( -\frac{v^{2}}{2}\right) }{\sqrt{2\pi }} \,
dv \\ &=-\frac{2}{\sqrt{2\pi \tau}}\nu \left( \tau\right) = -
\frac{2}{\sqrt{\pi (e^{2t} - 1)}} \nu(t),\end{aligned}
\end{equation*}%
where $\left( \tau -\tau ^{\prime }\right) =u$, $u=1/v^{2}$ and we have used
the fact that%
\begin{equation*}
\int_{0}^{\infty }\left( 1-v^{2}\right) \frac{\exp \left( -\frac{v^{2}}{2}%
\right) }{\sqrt{2\pi }}\,dv=0.
\end{equation*}%
Further,%
\begin{equation*}
\begin{aligned} \mathbb{{L}}_{2}&=\lim_{\varepsilon \rightarrow
0}\int_{0}^{\tau }\left( 1- \frac{\varepsilon^2}{\tau - \tau'} \right)
\frac{\exp \left(- \frac{\varepsilon^2}{2(\tau - \tau')} + \frac{\varepsilon
(b(\tau) - b(\tau'))}{\tau - \tau'} \right) \left(\Xi(\tau, \tau')
\nu(\tau') - \nu(\tau) \right)}{\sqrt{2\pi \left( \tau -\tau ^{\prime
}\right) ^{3}}} d\tau ^{\prime } \\ &=\lim_{\varepsilon \rightarrow
0}\int_{0}^{\tau } \frac{\exp \left(- \frac{\varepsilon^2}{2(\tau - \tau')}
+ \frac{\varepsilon (b(\tau) - b(\tau'))}{\tau - \tau'} \right)
\left(\Xi(\tau, \tau') \nu(\tau') - \nu(\tau) \right)}{\sqrt{2\pi \left(
\tau -\tau ^{\prime }\right) ^{3}}} d\tau ^{\prime } \\ &=\int_{0}^{\tau }
\frac{ \left(\Xi(\tau, \tau') \nu(\tau') - \nu(\tau) \right)}{\sqrt{2\pi
\left( \tau -\tau ^{\prime }\right) ^{3}}} d\tau ^{\prime }\end{aligned}
\end{equation*}%
where we have dropped the higher order $\varepsilon ^{2}$ term in the
integral 
in the second line. $\mathbb{L}_{3}$ is computed as in \cite%
{tikhonov2013equations}, pp. 530-535 
\begin{equation*}
\begin{aligned} \mathbb{{L}}_{3}&=\lim_{\varepsilon \rightarrow
0}\int_{0}^{\tau }\frac{\varepsilon (b(\tau) - b(\tau'))}{\tau - \tau'}
\frac{\exp \left(- \frac{\varepsilon^2}{2(\tau - \tau')} + \frac{\varepsilon
(b(\tau) - b(\tau'))}{\tau - \tau'} \right) \Xi(\tau, \tau')}{\sqrt{2\pi
\left( \tau -\tau ^{\prime }\right) ^{3}}}\nu \left( \tau ^{\prime }\right)
d\tau ^{\prime } \\ &= b'(\tau) \Xi \left( \tau,\tau\right) \nu \left(
\tau\right) = b e^{-t} \nu(t) ,\end{aligned}
\end{equation*}%
and 
\begin{equation*}
\begin{aligned} \mathbb{{L}}_{4}&=\lim_{\varepsilon \rightarrow
0}\int_{0}^{\tau }\frac{(b(\tau) - b(\tau'))^2}{\tau - \tau'} \frac{\exp
\left(- \frac{\varepsilon^2}{2(\tau - \tau')} + \frac{\varepsilon (b(\tau) -
b(\tau'))}{\tau - \tau'} \right) \Xi(\tau, \tau')}{\sqrt{2\pi \left( \tau
-\tau ^{\prime }\right) ^{3}}}\nu \left( \tau ^{\prime }\right) d\tau
^{\prime } \\ &=\int_{0}^{\tau}\left(\frac{b(\tau) - b(\tau')}{\tau - \tau'}
\right)^{2}\frac{\Xi \left( \tau,\tau^{\prime }\right) }{\sqrt{2\pi \left(
\tau-\tau^{\prime }\right) }}\nu \left( \tau^{\prime }\right) \, d
\tau^{\prime }.\end{aligned}
\end{equation*}%
Combining all terms, we finally have 
\begin{eqnarray*}
g(t) &=&-\frac{\left( e^{t}b-z\right) \exp \left( -\frac{\left( e^{t}b-z\right)
^{2}}{\left( e^{2t}-1\right) }+2t\right) }{\sqrt{\pi \left( e^{2t}-1\right)
^{3}}}-\left( e^{t}b+\frac{e^{2t}}{\sqrt{\pi \left( e^{2t}-1\right) }}%
\right) \nu \left( t\right) \\
&&+\frac{1}{2}e^{2t}\int_{0}^{\tau }\frac{\left( 1-2b^{2}\frac{\left( \sqrt{%
2\tau +1}-\sqrt{2\tau ^{\prime }+1}\right) }{\left( \sqrt{2\tau +1}+\sqrt{%
2\tau ^{\prime }+1}\right) }\right) \exp \left( -b^{2}\frac{\left( \sqrt{%
2\tau +1}-\sqrt{2\tau ^{\prime }+1}\right) }{\left( \sqrt{2\tau +1}+\sqrt{%
2\tau ^{\prime }+1}\right) }\right) \nu \left( \tau ^{\prime }\right) -\nu
\left( \tau \right) }{\sqrt{2\pi \left( \tau -\tau ^{\prime }\right) ^{3}}}%
d\tau ^{\prime }.
\end{eqnarray*}

\bibliographystyle{apalike}
\bibliography{main}

\end{document}